\documentclass[lettersize,journal]{IEEEtran}

\usepackage{amsmath,amsfonts}
\usepackage{algorithmic}
\usepackage{algorithm}
\usepackage{array}
\usepackage[caption=false,font=normalsize]{subfig}
\usepackage{textcomp}
\usepackage{stfloats}
\usepackage{url}
\usepackage{verbatim}
\usepackage{graphicx}
\usepackage{cite}
\usepackage{xcolor}

\usepackage{amssymb}

\newtheorem{remark}{Remark}

\begin{document}

\title{Joint Beamforming, Energy Management, and Trajectory Optimization for Figure-Eight Loitering in Solar-Powered HAPS-Enabled ISAC Systems
}
\author{
Xue Zhang,  {\em Member, IEEE}, Bang Huang,  {\em Member, IEEE}, \\ and Mohamed-Slim Alouini, {\em Fellow, IEEE}
\thanks{The authors are with Computer, Electrical and Mathematical Sciences
and Engineering (CEMSE) Division, Department of Electrical and Computer
Engineering, King Abdullah University of Science and Technology (KAUST), Thuwal 23955-6900, Saudi Arabia. (e-mail: xue.zhang@kaust.edu.sa; bang.huang@kaust.edu.sa; slim.alouini@kaust.edu.sa). 
}
\vspace{-8mm}
}
\maketitle

\vspace{-0.8cm}

\begin{abstract}
Solar-powered high-altitude platform stations (HAPSs) provide a promising platform for integrated sensing and communication (ISAC) owing to their wide-area coverage and long-endurance operation. This paper proposes a solar-powered HAPS-enabled ISAC framework for sustainable day-night operation, where a figure-eight loitering architecture is adopted to provide persistent ISAC services over geographically separated regions while harvesting solar energy. A unified communication-sensing-energy model is developed by jointly characterizing solar energy harvesting, battery dynamics, propulsion power consumption, communication transmission, and synthetic aperture radar (SAR) imaging. Based on this model, coupled optimization problems are formulated for daytime operation (DTO) and nighttime operation (NTO), where the battery state bridges the two operational phases through a long-term energy budget. The proposed framework jointly optimizes communication, sensing, mobility, and energy management to maximize daytime communication performance while minimizing nighttime propulsion energy consumption. Efficient iterative algorithms are developed to solve the resulting non-convex optimization problems. Simulation results verify the effectiveness of the proposed communication-sensing-energy co-design and demonstrate that the proposed framework effectively supports sustainable day-night ISAC operation.
\end{abstract}

\begin{IEEEkeywords}
HAPS, ISAC, solar energy harvesting, energy management, beamforming, trajectory.
\end{IEEEkeywords}

\section{Introduction}
\label{sec_introdution}
High-altitude platform stations (HAPSs) have attracted growing attention in recent years due to their wide coverage, reliable line-of-sight (LoS) communication links, and quasi-stationary positions relative to the Earth \cite{abbasi2024haps,yahia2022haps}. Operating in the stratosphere, HAPSs serve as an important bridge between terrestrial and satellite communication systems, extending connectivity and improving network resilience \cite{kurt2021vision}. These platforms can support a wide range of applications, including broadband communications, environmental monitoring, and wide-area surveillance \cite{mohammed2011role}. Recent advances in solar energy harvesting technologies further enable long-endurance and energy-autonomous operation, where photovoltaic (PV) panels harvest solar energy during daytime and onboard batteries store excess energy for nighttime operation \cite{javed2023interdisciplinary}.

Meanwhile, the rapid growth of next-generation wireless networks and the explosive increase in connected devices have created an unprecedented demand for both spectrum and energy resources. To address these challenges, integrated sensing and communication (ISAC) has emerged as a promising paradigm that enables communication and sensing functionalities to share spectrum, hardware, and energy resources \cite{zhang2024joint,liu2020joint,benaya2025aerial,zhang2025ris}. Benefiting from their broad coverage, persistent operation, and favorable observation geometry, HAPSs provide a natural platform for supporting ISAC services. In particular, synthetic aperture radar (SAR) imaging can be efficiently integrated with HAPS systems to provide high-resolution remote sensing while maintaining communication coverage \cite{wang2014high}.

Motivated by these observations, this paper develops a solar-powered HAPS-enabled ISAC framework that jointly integrates communication, SAR imaging, trajectory control, and energy management. Unlike conventional unmanned aerial vehicle (UAV)-enabled ISAC systems, persistent day-night operation introduces strong coupling among communication performance, sensing quality, flight control, solar energy harvesting, and battery dynamics. Moreover, many practical ISAC missions require persistent coverage over two geographically separated service regions. In such scenarios, a conventional circular loitering trajectory either remains far from one of the regions or requires a significantly larger orbit to cover both. To overcome this limitation, a figure-eight loitering trajectory is adopted to periodically revisit both regions while maintaining continuous fixed-wing flight. The proposed trajectory is therefore particularly suitable for persistent dual-region missions, whereas conventional circular loitering remains preferable when the service regions largely overlap. Based on a unified communication-sensing-energy model, we jointly optimize beamforming, trajectory control, and energy management to explicitly characterize the long-term tradeoff among communication performance, sensing quality, and energy sustainability.

\subsection{Related Work}
Solar-powered HAPSs have attracted significant attention owing to their capability of achieving long-endurance operation through solar energy harvesting and onboard energy storage \cite{karapantazis2005broadband}. Existing studies have investigated energy-aware communication design, trajectory optimization, and resource allocation for solar-powered aerial platforms \cite{marriott2020trajectory,javed2023interdisciplinary,nauman2017system,azzahra2019noma,ji2020energy,hsieh2020uav}. For example, \cite{marriott2020trajectory} optimized aircraft trajectories to balance harvested and consumed energy, while \cite{javed2023interdisciplinary} developed an interdisciplinary framework integrating aerodynamics, communications, and renewable energy management. Nevertheless, these studies primarily focus on communication and energy management without explicitly considering integrated sensing functionalities.

The emergence of integrated sensing and communication (ISAC) has motivated extensive research on the joint optimization of communication and sensing performance for aerial platforms. Existing works have investigated communication-sensing beamforming design, including energy-efficient beamforming and distributed/bistatic multiple-input multiple-output (MIMO) ISAC architectures \cite{he2025energy,he2025bistatic}. Furthermore, the mobility of aerial platforms provides an additional degree of freedom for improving ISAC performance, leading to extensive studies on the joint design of trajectory optimization, beamforming, and resource allocation for UAV-enabled ISAC systems \cite{khalili2023energy,khalili2024efficient,li2025control,deng2024joint,ning2026joint}. Representative examples include the energy-aware UAV-ISAC framework in \cite{khalili2023energy}, the limited-backhaul UAV-enabled ISAC framework in \cite{khalili2024efficient}, the control-based trajectory optimization approach in \cite{li2025control}, and the joint beamforming and trajectory optimization schemes proposed in \cite{deng2024joint,ning2026joint}. These studies demonstrate the benefits of mobility-aware communication-sensing co-design. Recent work in \cite{lyu2025empowering} further emphasized the importance of integrating sensing, communication, and computation for resource-efficient aerial networks.

Compared with low-altitude UAV platforms, HAPS-enabled ISAC systems operate in the stratosphere and are expected to provide persistent wide-area coverage over extended periods. Consequently, renewable energy harvesting, battery dynamics, and propulsion energy consumption become fundamental design considerations. Recent studies have begun extending ISAC to HAPS platforms through joint communication-sensing optimization \cite{kanani2025optimizing,zhang2026design}. For example, \cite{kanani2025optimizing} investigated a HAPS-UAV integrated ISAC architecture, while \cite{zhang2026design} jointly optimized HAPS deployment and beamforming under SAR imaging constraints. In parallel, joint communication and SAR imaging (JCASAR) has emerged as a promising paradigm for enabling communication and imaging systems to share spectrum, hardware, and energy resources. Existing studies have mainly focused on waveform design and dual-function radar-communication architectures~\cite{wang2019first,tan2022joint,yang2022waveform,liu2022integrated,zheng2024waveform}, demonstrating the feasibility of integrating communication and SAR imaging.

Despite these advances, existing studies generally optimize communication, sensing, trajectory control, or energy management over a single operational period. The long-term coupling among communication performance, SAR imaging, mobility, renewable energy harvesting, battery dynamics, and day-night operation has not yet been systematically investigated. This gap motivates the unified communication-sensing-energy co-design framework proposed in this paper, which jointly integrates communication transmission, SAR imaging, trajectory control, solar energy harvesting, and battery energy management to enable sustainable day-night operation for solar-powered HAPS-enabled ISAC systems.

\subsection{Main Contributions} 
This paper develops a solar-powered HAPS-enabled ISAC framework that jointly integrates communication, SAR imaging, figure-eight loitering, and energy management for sustainable day-night operation. The main contributions are summarized as follows.
\begin{itemize}
\item We propose a figure-eight loitering architecture for solar-powered HAPS-enabled ISAC systems to provide persistent ISAC services over two geographically separated service regions while maintaining continuous fixed-wing flight. 
\item We develop a unified communication-sensing-energy model that jointly characterizes solar energy harvesting, battery dynamics, propulsion power consumption under steady circular flight (SCF), communication transmission, and SAR imaging, thereby characterizing the coupling among communication, sensing, mobility, and energy sustainability. 
\item We formulate coupled optimization problems for daytime operation (DTO) and nighttime operation (NTO), where the battery state explicitly links the two operational phases through a long-term energy budget. 
\item During DTO, the flight speed profile, transmit beamforming, and sensing covariance matrix are jointly optimized to maximize the communication sum rate subject to sensing, mobility, power, and energy constraints. During NTO, the flight speed and altitude are jointly optimized to minimize propulsion energy consumption under the available battery energy. Efficient iterative algorithms are developed to solve the resulting non-convex optimization problems. 
\item Numerical results demonstrate that, for the considered dual-region deployment scenario, the proposed framework achieves a higher communication sum rate than the benchmark trajectories while satisfying the prescribed sensing and energy constraints. Moreover, the proposed NTO strategy significantly reduces propulsion energy consumption, thereby supporting sustainable day-night operation. \end{itemize}

\subsection{Outline and Notations}
\textit{Outline}: The remainder of the paper is organized as follows. Section~\ref{sec_model} introduces the solar-powered HAPS-enabled ISAC system model. Section~\ref{sec_problem} formulates two optimization problems for the DTO and NTO phases. Sections~\ref{sec_method_DTO} and~\ref{sec_method_NTO} develop efficient solution methods for the DTO and NTO problems, respectively. Numerical results are presented in Section~\ref{sec_sim} to evaluate the performance of the proposed framework. Finally, Section~\ref{sec_con} concludes the paper.

\textit{Notations}: The vectors (matrices) are denoted by lower-case (upper-case) boldface characters. $\mathbf{I}_{M}$ stands for the $M\times M$ identity matrix, and $\mathbf{0}$ represents a zero matrix with appropriate dimension. $\dot{\mathbf{x}}(t)$ denotes the derivative of the time-dependent function $\mathbf{x}(t)$ with respect to the time $t$. Superscripts $T$, and $H$ represent the transpose and conjugate transpose, respectively. $\mathbb{E}\{\cdot\}$ represents the expectation operator. The trace of $\mathbf{A}$ are represented by $\mathrm{tr}(\mathbf{A})$. $[\mathbf{A}]_{i,j}$ stands for the $(i,j)$ entry of $\mathbf{A}$. $\mathbf{A}\succeq\mathbf{0}$ is equivalent to $\mathbf{A}$ being positive semidefinite. $|\cdot|$ stand for the magnitude of a complex number, and $\|\cdot\|$ is the Euclidean norm of a complex vector. $\mathcal{O}(\cdot)$ denotes the convolution Big-O notation.

\par

\begin{figure}[t]
\centering
\includegraphics[width=1.0\linewidth]{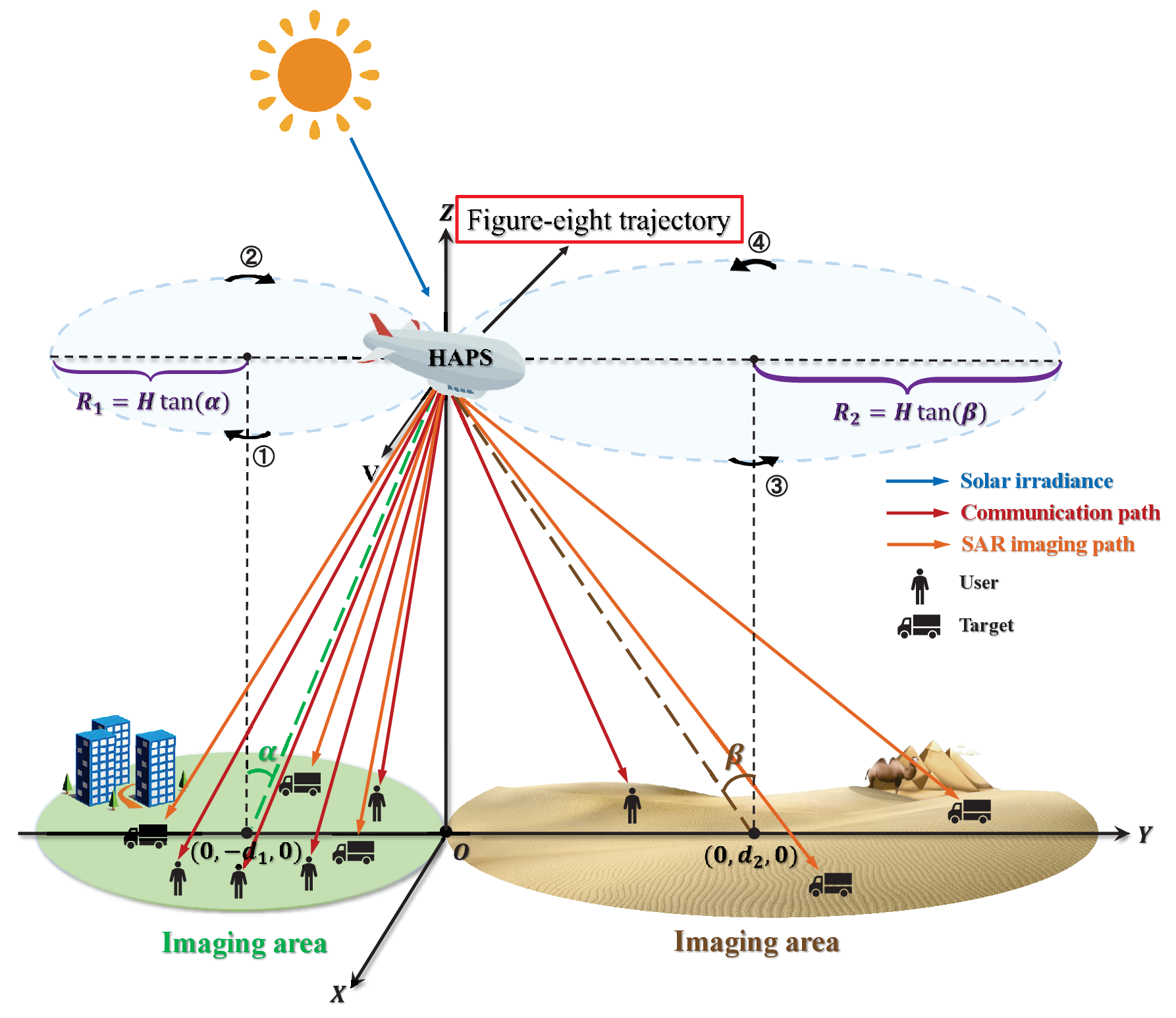}
\caption{Solar-powered HAPS-enabled ISAC system model with the figure-eight trajectory.}
\label{system_model}
\end{figure}

\section{System Model}
\label{sec_model} 
As illustrated in Fig.~\ref{system_model}, we consider a fixed-wing solar-powered HAPS equipped with an ISAC payload, employing circular SAR~\cite{soumekh1996reconnaissance} for high-resolution imaging of $Q \ge 1$ ground targets while providing downlink communication to $K \ge 1$ single-antenna ground users. The HAPS employs an $M$-element vertical uniform linear array (ULA) with half-wavelength spacing $d_s=\lambda/2$, where $\lambda$ denotes the carrier wavelength. The ULA is rigidly mounted with its axis aligned with the global $Z$-axis, assuming a constant HAPS attitude throughout each DTO flight period. In a three-dimensional (3D) Cartesian coordinate system, the HAPS follows a figure-eight trajectory at altitude $H$, consisting of two circular segments with radii $R_1=H\tan(\alpha)$ and $R_2=H\tan(\beta)$, where $\alpha,\beta\in(0,\pi/2)$ denote the SAR observation angles. The two segments are centered at $(0,-R_1,H)$ and $(0,R_2,H)$, respectively, and intersect at $(0,0,H)$.

Starting from the intersection point, the HAPS traverses the left and right circular segments with tangential velocities $V_1(t)$ and $V_2(t)$ over $[0,T_1]$ and $[T_1,T]$, respectively, where $T$ denotes the full trajectory period. This closed trajectory enables 
periodic SAR imaging and continuous downlink communication while providing angular diversity. Specifically, the horizontal position of the HAPS is given by
\begin{align}
\mathbf{p}(t)=
\begin{cases}
[R_1\sin(\varphi_1(t)),  R_1(\cos(\varphi_1(t))-1)],\,\, 0 \le t \le T_1,\\[2mm]
[R_2\sin(\varphi_2(t)),R_2(1-\cos(\varphi_2(t)))],\,\,T_1 \le t \le T,
\end{cases} \notag
\end{align}
where $\varphi_1(t)=\int_0^t \frac{V_1(\tau)}{R_1}d\tau$ and $\varphi_2(t)=\int_{T_1}^t \frac{V_2(\tau)}{R_2}d\tau$ denote the angular displacements along the two segments, respectively. To ensure trajectory closure, $\varphi_1(T_1)=\varphi_2(T)=2\pi$. The flight speed is constrained by $V_{\mathrm{min}}\leq\|\dot{\mathbf{p}}(t)\|\leq V_{\mathrm{max}}$.

For tractability, the figure-eight trajectory period $T$ is divided into $N$ equal-duration time slots indexed by $n=1,\ldots,N$, with $\Delta_t=T/N$. Within each sufficiently short slot, the HAPS position is assumed constant. The predetermined trajectory is parameterized by two angular variables, $\varphi_1[n]$ and $\varphi_2[n]$, governed by the flight speed profile. Accordingly, the horizontal position of the HAPS is given by~\cite{hu2022trajectory,hua20203d,zeng2017energy}
\begin{equation}
\begin{aligned}
&\mathbf{p}[n]=\\
&\begin{cases}
R_1[\sin(\varphi_1[n]),\cos(\varphi_1[n])-1], n=1,\ldots,N_1+1,\\[2mm]
R_2[\sin(\varphi_2[n]),1-\cos(\varphi_2[n])], n=N_1+1,\ldots,N+1.
\end{cases}
\end{aligned}
\label{trajectory_cons1}
\end{equation}
where the angular variables are initialized as
\begin{align}
\begin{array}{ll}
    \varphi_1[1]=0, &  \varphi_1[N_1+1]=2\pi\\
    \varphi_2[N_1+1]=0, & \varphi_2[N+1]=2\pi,
\end{array}
\end{align}
and evolve according to
\begin{align}
    &\varphi_1[n+1]=\varphi_1[n]+\frac{V[n]\Delta_t}{R_1},
    \quad n=1,\ldots,N_1,\\
    &\varphi_2[n+1]=\varphi_2[n]+\frac{V[n]\Delta_t}{R_2},
    \quad n=N_1+1,\ldots,N.
\end{align}
The figure-eight trajectory satisfies the closed-loop condition $\mathbf{p}[1]=\mathbf{p}[N_1+1]=\mathbf{p}[N+1]$, while the flight speed is constrained as
\begin{align}
    V_{\mathrm{min}}\leq V[n]\leq V_{\mathrm{max}},\quad \forall n, \label{trajectory_cons3}
\end{align}
where $V_{\mathrm{min}}$ and $V_{\mathrm{max}}$ denote the minimum and maximum allowable flight speeds, respectively.

Since the HAPS simultaneously transmits SAR imaging signals and downlink communication signals, the transmitted signal at time slot $n$ is expressed as
\begin{align}\label{transmitted signal}
    \mathbf{x}[n] = \sum_{k=1}^K\mathbf{w}_k[n]s_k[n] + \mathbf{s}_{0}[n],  
\end{align}
where $\mathbf{s}_0[n]\in\mathbb{C}^{M\times 1}$ denotes the dedicated SAR imaging signal at time slot $n$, modeled as a zero-mean independent random vector with covariance matrix $\mathbf{R}_s[n]=\mathbb{E}\{\mathbf{s}_0[n]\mathbf{s}_0^H[n]\}\succeq \pmb{0}$, $s_k[n]\sim\mathcal{CN}(0,1)$ represents the communication signal intended by the $k$th user, described as zero-mean, unit-variance circularly symmetric complex Gaussian random variable, and $\mathbf{w}_k[n]\in\mathbb{C}^{M\times 1}$ is the corresponding transmit beamforming vector. Accordingly, the average transmit power at slot $n$ is given by  
\begin{align}\label{average transmit power}
    P_{\mathrm{ave}}[n] =& \mathbb{E}\{\|\mathbf{x}[n]\|^2\} = \sum_{k=1}^K\|\mathbf{w}_k[n]\|^2 + \mathrm{tr}(\mathbf{R}_s[n]). 
\end{align}
The transmit power satisfies $P_{\mathrm{ave}}[n]\leq P_{\mathrm{max}}$, 
where $P_{\mathrm{max}}$ is the maximum allowable transmit power.

\subsection{Channel Model and Performance Metrics}
Owing to the high operating altitude of the HAPS, a strong LoS link typically exists between the platform and each ground user. However, in urban or partially obstructed environments, scattering and reflection may contribute additional non-line-of-sight (NLoS) components. To account for both effects, the downlink channel is modeled as a 3D Rician fading channel composed of a dominant LoS component and a diffuse NLoS component. Accordingly, the channel vector between the HAPS and the $k$th user at time slot $n$ is expressed as~\cite{abbasi2024hemispherical,shamsabadi2024enhancing}
\begin{align}
    &\mathbf{g}(\mathbf{p}[n],\mathbf{u}_k)=\frac{1}{\sqrt{\ell}} \notag\\
    &\quad\times\Big(\sqrt{\frac{K_u}{K_u+1}}\mathbf{g}_{k,\mathrm{LOS}}[n]+\sqrt{\frac{1}{K_u+1}}\mathbf{g}_{k,\mathrm{NLOS}}[n]\Big),
\end{align}
where $K_u$ is the Rician factor,
$\ell=d^2(\mathbf{p}[n],\mathbf{u}_k)/\rho_0$
denotes the free-space path loss,
with $\rho_0=\left(\frac{\lambda}{4\pi}\right)^2$ representing the reference channel power gain at a distance of 1 m,
and $d(\mathbf{p}[n],\mathbf{u}_k)
=\sqrt{\|\mathbf{p}[n]-\mathbf{u}_k\|^2+H^2}$
denotes the distance between the HAPS and
the $k$th user at time slot $n$. $\mathbf{g}_{k,\mathrm{NLOS}}[n]\sim\mathcal{CN}(\pmb{0},\mathbf{I}_M)$ represents the NLOS component, whose elements are independently drawn from a complex Gaussian distribution with zero mean and unit variance, and $\mathbf{g}_{k,\mathrm{LOS}}[n]=\mathbf{a}_k(\mathbf{p}[n],\mathbf{u}_k)$ denotes the LoS component, represented by the steering vector associated with the $k$th user. Owing to the adopted ULA deployment and constant HAPS attitude, the steering vector depends only on the instantaneous relative geometry between the HAPS and the user, and is given by
\begin{align}\label{steering vector}
    &\mathbf{a}_k(\mathbf{p}[n],\mathbf{u}_k)=[1,\ldots,e^{j\pi(M-1)\cos(\theta(\mathbf{p}[n],\mathbf{u}_k))}]^T. 
\end{align}
Here, $\theta(\mathbf{p}[n],\mathbf{u}_k)$ is the angle of departure (AoD) towards the $k$th user, given by 
\begin{align}
    \theta(\mathbf{p}[n],\mathbf{u}_k) = \arccos\Big(\frac{H}{\sqrt{\|\mathbf{p}[n]-\mathbf{u}_k\|^2+H^2}}\Big).
\end{align}
The vertical ULA is adopted for tractable optimization rather than array design. Although users with identical AoDs share the same LoS steering vectors, HAPS mobility may alleviate this 
angular ambiguity. The proposed framework can be extended to uniform planar arrays (UPAs) with appropriate modifications to the channel model and optimization procedure. Accordingly, the received signal at the $k$th user during time slot $n$ can be expressed as
\begin{align}\label{received signal}
    y_k[n] = \mathbf{g}_k^H(\mathbf{p}[n],\mathbf{u}_k)\mathbf{x}[n] + n_k[n],
\end{align}
where $n_k[n]\sim\mathcal{CN}(0,\sigma_k^2)$ denotes the additive white Gaussian noise with noise power $\sigma_k^2$.

\begin{figure*}[ht!]
\begin{align}\label{gamma_k}
    \mathrm{SINR}_k(\mathbf{p}[n],\{\mathbf{w}_k[n]\},\pmb{\mathcal{R}}_s[n]) = \frac{\mathbb{E}\{|\mathbf{g}_k^H(\mathbf{h}[n],\mathbf{m}_k)\mathbf{w}_k[n]|^2\}}{\mathbb{E}\Big\{\sum\limits_{p=1\atop p\neq k}^K|\mathbf{g}_k^H(\mathbf{h}[n],\mathbf{m}_k)\mathbf{w}_p[n]|^2\Big\}+\mathbb{E}\{|\mathbf{g}_k^H(\mathbf{h}[n],\mathbf{m}_k)\mathbf{s}_0[n]|^2\}+\sigma_k^2}.
\end{align}
\hrulefill
\end{figure*}

\subsubsection{Communication metric}
From (\ref{received signal}), each user experiences multiuser interference from the communication signals $s_i[n]$ ($p\neq k$) intended for other users, as well as interference introduced by the sensing signal $\mathbf{s}_0[n]$. The resulting SINR at the $k$th user during time slot $n$ can be expressed as (\ref{gamma_k}). Under a Gaussian noise environment, the achievable rate of the $k$th user at time slot $n$, measured in bits-per-second-per-Hertz (bps/Hz), is given by 
\begin{align}
    \mathcal{R}_k[n] = B_c\log_2(1+\mathrm{SINR}_k(\mathbf{p}[n],\{\mathbf{w}_k[n]\},\mathbf{R}_s[n]),  
\end{align}
where $B_c=200$ MHz is the communication bandwidth.  

\subsubsection{SAR imaging metric}
When a finite set of $Q$ points is considered and the existence or exact locations of potential targets are unknown, these points are uniformly sampled across the entire region of interest to represent possible target positions. Alternatively, when the HAPS employs the SAR-GMTI technique~\cite{li2011influence,makhoul2014performance} for target tracking and approximate target locations are available, the points $\mathbf{t}_q$ can be assigned to these estimated positions. The SAR imaging process jointly exploits the dedicated sensing signal $\mathbf{s}_0$ and the communication signals $s_k$ for $k=1, \ldots, K$. The imaging performance is then characterized by the transmit beam pattern gains at the $Q$ potential target points, defined as
\begin{align}
    &B(\mathbf{p}[n],\mathbf{t}_q)=\mathbb{E}\{|\mathbf{a}^H(\mathbf{p}[n],\mathbf{t}_q)\mathbf{x}[n]|^2\} \notag \\
    &= \mathbf{a}^H(\mathbf{p}[n],\mathbf{t}_q)\Big(\sum_{k=1}^K\mathbf{w}_k[n]\mathbf{w}_k^H[n]+\mathbf{R}_s[n]\Big)\mathbf{a}(\mathbf{p}[n],\mathbf{t}_q), \notag
\end{align}
where $\mathbf{a}(\mathbf{p}[n],\mathbf{t}_q)$ is the steering vector towards the $q$th potential target as defined in (\ref{steering vector}).

Moreover, the received SNR for SAR imaging at time slot $n$ can be written as~\cite{curlander1991synthetic,kim2009antenna,lahmeri2022trajectory}
\begin{align}\label{SNR}
    \mathrm{SNR}_{\psi}[n] = \frac{P_{\mathrm{ave}}[n]C_{\mathrm{SAR}}\sin^2(\psi)}{V[n]}, \quad \psi\in\{\alpha,\beta\},
\end{align}
Here, $C_{\mathrm{SAR}} = \frac{G_tG_r\lambda^3\sigma_0c\tau_p\mathrm{PRF}}{256\pi^3H^3\kappa T_oNFB_rL_{tot}}$, $G_t=35$\,dBi and $G_r=35$\,dBi denote the radar antenna gains for transmission and reception, respectively. $\sigma_0=1$ is the backscattering coefficient, $c=3\times 10^{8}$ m/s is the speed of light, $\tau_p=10$ $\mu$s represents the radar pulse duration, $\mathrm{PRF}=2$ kHz denotes the radar pulse repetition frequency, $\kappa=1.380649\times10^{-23}$ J/K is the Boltzmann's constant, $B_r=200$ MHz represents the radar bandwidth, $L_{tot}=10$ dB is the combined losses, $NF=6$\,dB stands for the system noise figure, and $T_o=290$ K denotes the equivalent noise temperature.

\par

\begin{figure}[t]
\centering
\includegraphics[width=0.4\linewidth]{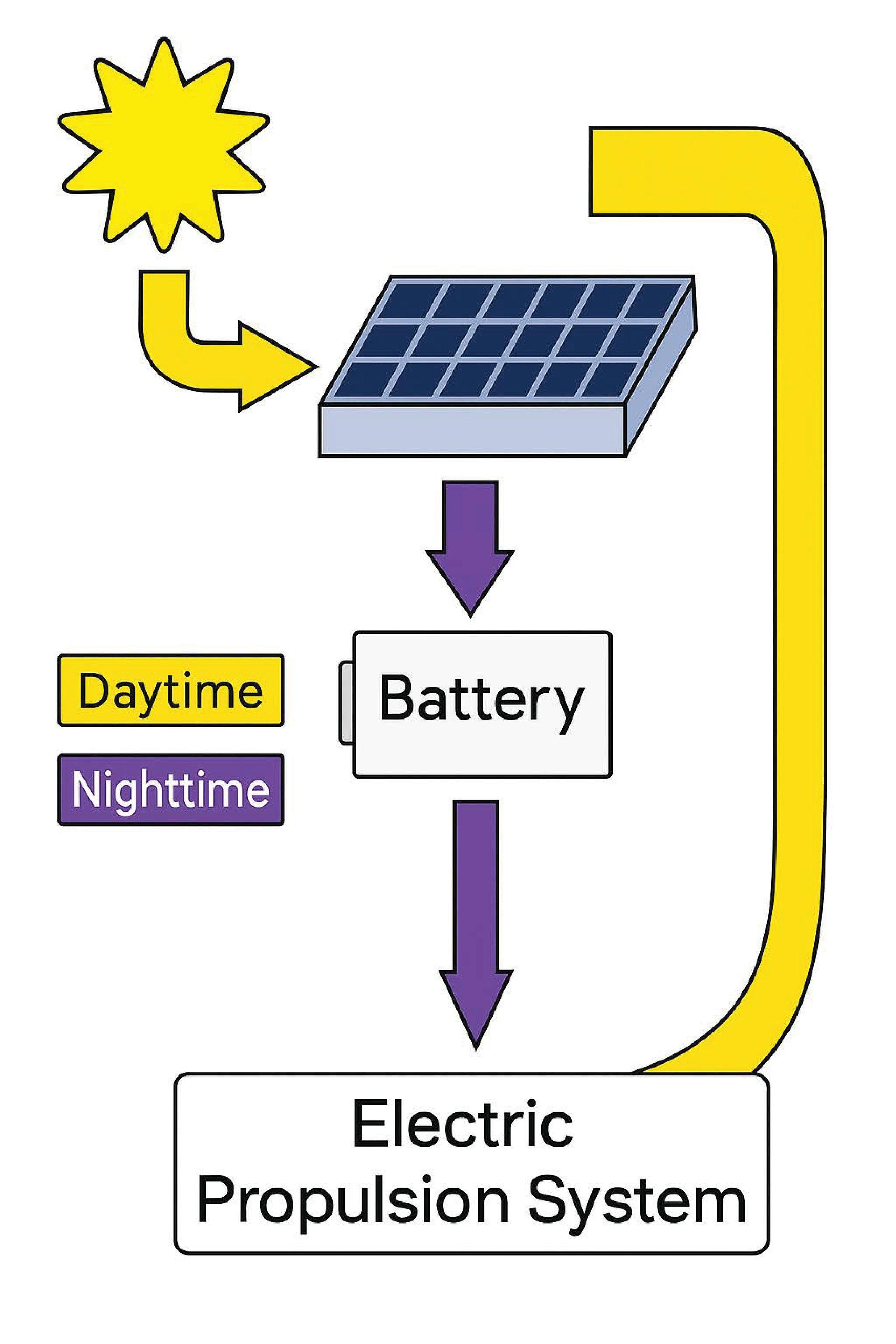}
\caption{Energy flow diagram of the solar-powered HAPS during the DTO and NTO phases.}
\label{solar_energy}
\end{figure}

\subsection{Solar Energy Model}
The HAPS employs horizontal solar panels mounted on its upper surface. As illustrated in Fig.~\ref{solar_energy}, during DTO, the harvested solar energy supports station-keeping and communication, while the surplus is stored in rechargeable batteries for NTO. To ensure energy-neutral operation over a 24-hour cycle, we model daytime energy collection, nighttime energy consumption, and battery dynamics for subsequent trajectory and energy optimization.

\subsubsection{DTO - Energy collection}
The long-endurance HAPS relies entirely on harvested solar energy to sustain flight and ISAC operations, necessitating an accurate solar irradiance model. Solar irradiance, measured in $\mathrm{W}/\mathrm{m}^2$, quantifies the radiant power received per unit area. For a HAPS at latitude $\chi$ and altitude $H$, the solar irradiance on a given date, accounting for Earth's orbital eccentricity and atmospheric absorption, is expressed as~\cite{bolandhemmat2019energy}
\begin{align}\label{solar_irridiance_hours}
    I_h(H,t_h) = I_0\Big(1+0.034\cos\Big(\frac{2\pi j_d}{365}\Big)\Big)f(H,t_h), 
\end{align}
where $t_h$ is the hours of the given date, $I_0=1366.1$\,$\mathrm{W}/\mathrm{m}^2$ is the standard solar constant at zero air mass defined by American Society for Testing and Materials (ASTM E490), $j_d$ denotes the Julian day corresponding to the given date, and $f(H,t_h)$ represents the atmospheric absorption factor given by~\cite{aglietti2009harnessing} 
\begin{align}
    f(H,t_h)= \mathrm{exp}\left(-P_R(H)m_R(\epsilon_s(t_h))\alpha_{\mathrm{ext}}\right),
\end{align}
with $\alpha_{\mathrm{ext}}=0.32$ being the extinction coefficient under clear-sky conditions~\cite{aglietti2009harnessing}. $P_R(H)$ denotes the relative atmospheric pressure, calculated according to the International Standard Atmosphere (ISA) and the 1976 U.S. Standard Atmosphere~\cite{atmosphere1976us}, expressed as
\begin{align}
    P_R(H) = \left\{\begin{array}{ll}
        \frac{P_{b_1}}{P_0}\mathrm{exp}\Big(\frac{-gM(H-11)}{RT_b}\Big) & H\in[11,20]\,\mathrm{km}, \\
        \frac{P_{b_2}}{P_0}\left(\frac{T_b}{T_b+L_b(H-20)}\right)^{\frac{gM}{RL_b}}, & H\in[20,32]\,\mathrm{km},
    \end{array}\right. \notag
\end{align}
where $P_0=101325$ Pa is the atmosphere pressure at mean sea level, $P_{b_1}=22632.06$ Pa, and $P_{b_2} =
5474.889$ Pa are the base static pressures corresponding to the base altitudes $11$ km and $20$ km, respectively. $g=9.8$ $\mathrm{m/s}^2$ is the gravitational acceleration, and $M=0.0289644$ kg/mol is the molar mass of Earth’s air,  $R = 8.31432$ N · m/mol · K is the universal gas constant, and $T_b = 216.65$ K is the base temperature with base temperature lapse rate $L_b$ = $1$ K/km.  $m_R(\epsilon_s(t_h))$ is the relative air mass given by~\cite{reda2004solar} 
\begin{align}
    m_R(\epsilon_s(t_h)) = \frac{1}{\cos(90^{\circ}-\epsilon_s(t_h))},
\end{align}
where $\epsilon_s(t_h)$ denotes the solar elevation angle, calculated as $\epsilon_s(t_h) = \arcsin(\sin(\chi)\sin(\delta)+\cos(\chi)\cos(\delta)\cos(H_a(t_h)))$ with $H_a(t_h)=15(t_h-12)$ being the solar hour angle, and $\delta = 23.45^{\circ}\sin(360 (284 + j_d) / 365)$ representing the solar declination angle.

The instantaneous solar irradiance $I_h(H,t_h)$ in (\ref{solar_irridiance_hours}) characterizes the daily variation of solar irradiance at a given altitude and date. Since $I_h(H,t_h)$ is negligible at night, the total incident solar energy per unit area is obtained by integrating the irradiance over the daylight period. For the considered fixed-wing stratospheric HAPS operating in quasi-stationary SCF, attitude variations are assumed to be sufficiently small during normal operation. Accordingly, the solar panels are approximated as horizontal, capturing the dominant effects of solar elevation and atmospheric absorption while maintaining analytical tractability. By Lambert's cosine law, the irradiance incident on a surface is proportional to the cosine of the angle between the incident solar rays and the surface normal. Hence, the instantaneous solar irradiance incident on the horizontal solar panels is given by
\begin{align}
I_i(H,t_h) = I_m(H,t_h)\cos(90^{\circ}-\epsilon_s(t_h)),
\end{align}
where $I_m(H,t_h)$ denotes the maximum solar irradiance at solar noon on a given day, obtained from~(\ref{solar_irridiance_hours}). Accordingly, the total irradiance per unit area on day $t_d$, denoted as $I_d(H,t_d)$, is obtained by integrating the instantaneous irradiance $I_i(H,t_h)$ over the sunlight duration. It can be further estimated using an approximate elevation-angle time series model as follows:\cite{arum2020energy}
\begin{align}
    I_d(H,t_d) = \frac{I_m(H,t_d)\tau(t_d)}{\epsilon_{\mathrm{sm}}(t_d)}(1-\cos(\epsilon_{\mathrm{sm}}(t_d))), 
\end{align}
where $I_m(H,t_d)$ denotes the maximum solar irradiance at solar noon on day $t_d$, obtained from (\ref{solar_irridiance_hours}), $\epsilon_{\mathrm{sm}}$ represents the maximum solar elevation angle on a day $t_d$, which can be estimated as $\epsilon_{\mathrm{sm}} = 90^{\circ} - |\chi - \delta|$. The daylight time duration $\tau(t_d)$ is expressed as
\begin{align}
    \tau(t_d) = 24\left(1-\frac{1}{\pi}\cos^{-1}\left(\frac{\tan(\chi)\sin(\varepsilon_0)\sin(\Phi(t_d))}{\sqrt{1-\sin^2(\varepsilon_0)\sin^2(\Phi(t_d))}}\right)\right), \notag
\end{align}
where $\varepsilon_0=23.44^{\circ}$ is the Earth’s axial tilt (obliquity) with respect to the ecliptic plane, $M_a(t_d)=-0.041+0.017202t_d$ denotes the mean angular distance (mean anomaly) of the Sun, and $\Phi(t_d)$ is the solar azimuth angle, given by
\begin{align}
    \Phi(t_d) =& -1.3411 + M_a(t_d) + 0.0334\sin(M_a(t_d)) \notag\\
            &+ 0.0003\sin(2M_a(t_d)). 
\end{align}

As a result, the total available solar energy harvested by the PV cells on a given day, expressed in watt-hours, can be written as 
\begin{align}
    E_s(H,t_d) = \eta_sA_pI_d(H,t_d),
\end{align}
where $\eta_s$ denotes the power conversion efficiency of solar panels, assumed to be 0.29, and $A_p$ represents the total effective area of the solar panels installed on the HAPS platform~\cite{javed2023interdisciplinary}. Similarly, the instantaneous available solar power at any given time during the day is expressed as 
\begin{align}
P_a(H,t_h) = \eta_s A_p I_h(H,t_h).
\end{align}

\begin{figure}[t]
\centering
\includegraphics[width=0.75\linewidth]{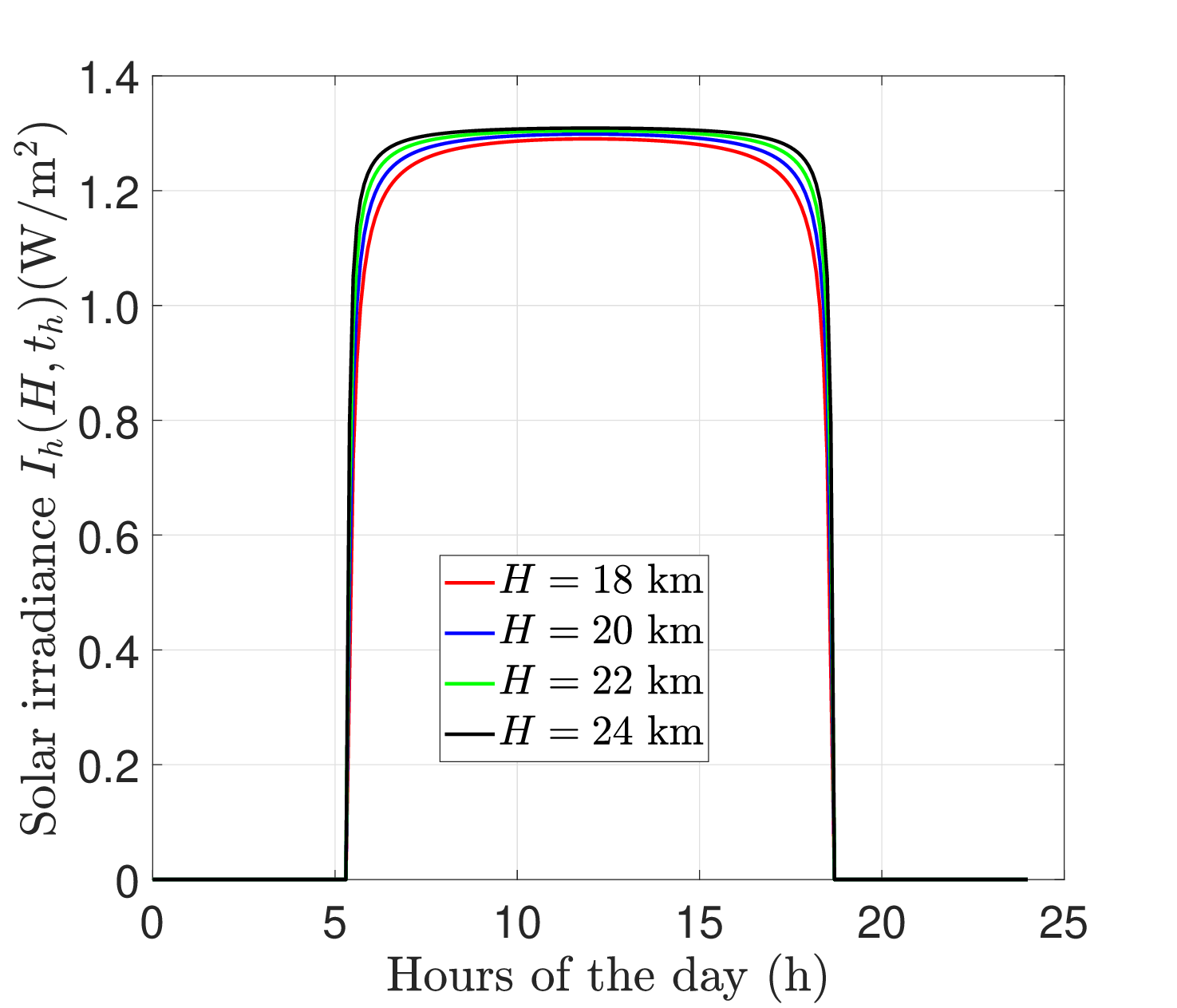}
\caption{Solar irradiance versus hour of the day under different flight altitudes.}
\label{It_hours}
\end{figure}

\begin{figure}[t]
\centering
\includegraphics[width=0.75\linewidth]{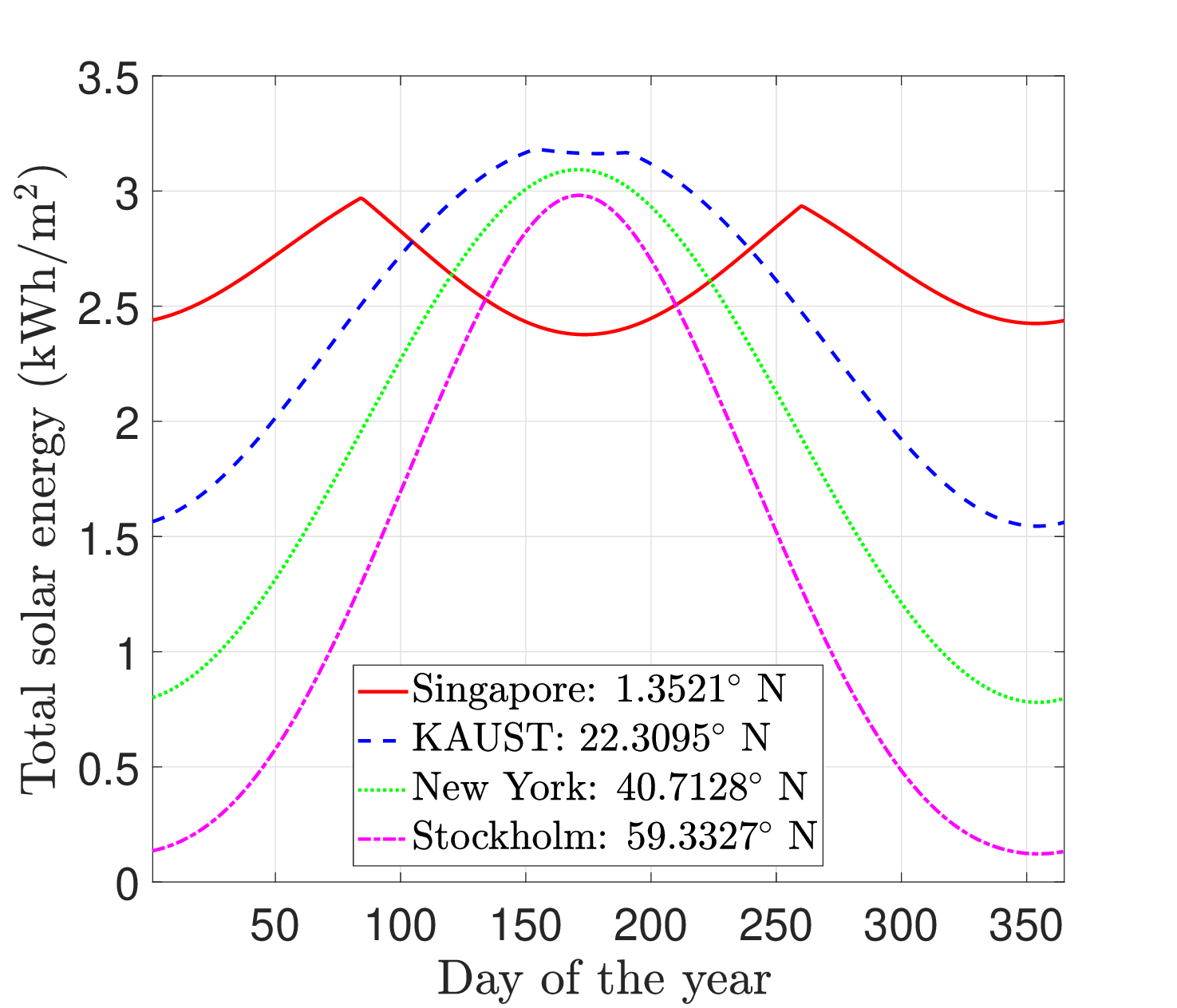}
\caption{Annual variation of total daily solar energy per unit area for different latitudes.}
\label{E_days}
\end{figure}

\begin{remark}
As an illustrative example, we consider a HAPS located at the latitude of King Abdullah University of Science and Technology (KAUST), $\chi=22.3097^\circ$. Fig.~\ref{It_hours} illustrates the instantaneous solar irradiance on June~21, 2024 (Julian day $j_d=173$) at different flight altitudes. The irradiance exhibits only a slight increase with altitude due to reduced atmospheric attenuation, indicating that altitude has a limited impact on the harvested solar energy. Moreover, owing to the quasi-stationary flight of the fixed-wing HAPS and the assumed constant platform attitude, the incident solar flux is primarily determined by the solar elevation angle, making the effect of horizontal mobility negligible. Fig.~\ref{E_days} further shows the annual variation of the total daily solar energy per unit area at different latitudes. Unlike altitude, latitude significantly affects the harvested solar energy through seasonal variations in daylight duration and solar elevation angle, with larger seasonal fluctuations observed at higher latitudes. Therefore, latitude is the dominant factor governing long-term solar energy availability. The adopted harvesting model assumes representative clear-sky stratospheric conditions, while incorporating detailed attitude dynamics and stochastic weather effects is left for future work.
\end{remark}

\par

\subsubsection{NTO - Energy consumption}
Building upon the DTO energy collection model, we now turn to the NTO phase, during which the energy stored during the daytime is primarily consumed for propulsion, while additional power is allocated to SAR imaging and communication. 
To maintain station-keeping at a fixed altitude during the NTO phase, the HAPS must continuously generate sufficient propulsion power to sustain SCF. The SCF is typically achieved through steady horizontal flight (SHF) with a constant bank angle $\vartheta$, during which the vertical component of the lift force $F_{lv}$ balances the HAPS weight $F_w$. In this condition, the lift force $F_l$ counteracts the gravity, while the thrust $F_t$ compensates for the drag force $F_d$, ensuring stable horizontal motion. Following the aerodynamic principles described in~\cite{stengel2004flight,arum2020energy,javed2023interdisciplinary}, the propulsion power required for SCF at time slot $n$ can be expressed as 
\begin{align}\label{P_SCF}
    P_{\mathrm{SCF}}[n] =& \frac{1}{\cos(\vartheta)^2}P_{\mathrm{SHF}}[n], 
\end{align}
where $P_{\mathrm{SHF}}[n]$ represents the propulsion power under SHF conditions and is given by $P_{\mathrm{SHF}}[n] = F_t[n]V[n]/(f_p f_e)$ with $f_p=0.85$ and $f_e=0.90$ being the propeller and engine efficiencies, respectively. The corresponding thrust $F_t[n]$ required to overcome aerodynamic drag at time slot $n$ is modeled as
\begin{align}
    F_t[n] = \frac{1}{2}\rho_hV[n]^2SC_{D_0} + \frac{2\epsilon F_w^2}{\rho_hSV[n]^2}. 
\end{align}
Here, $\rho_h$ represents the air density, approximated for the relevant altitude range by the curve-fitting expression $(0.95162H^2-52.29356H+753.39927)\times 10^{-3}$\,$kg/m^3$, $S=143$ m$^2$ denotes wing surface area, $C_{D_0}=0.015$ is the zero-lift drag coefficient, and $\epsilon=(\pi e_oAR_w)^{-1}$ with Oswald's efficiency factor $e_o=0.6385$ and wing aspect ratio $AR_w=30$. 

\subsubsection{Energy storage}
Building upon the DTO and NTO models, the energy storage process characterizes how the HAPS balances power inflow and outflow to sustain continuous operation. During the DTO, the excess solar power remaining after supporting communication and propulsion is stored in the onboard rechargeable batteries. Conversely, during the NTO, the stored energy is discharged to supply the power required for propulsion, SAR imaging, and communication. In particular, during the DTO phase, the instantaneous net power balance at time slot $n$ determines whether the battery is charging or discharging, and is expressed as
\begin{align}
    P_{\mathrm{net}}[n] = P_a - P_{\mathrm{SCF}}[n] - P_{\mathrm{ave}}[n], 
\end{align}
where $P_a$ denotes the available solar power, $P_{\mathrm{SCF}}$ is the propulsion power required for SCF, and $P_{\mathrm{ave}}[n]$ represents the total power allocated to communication and sensing. The term $P_{\mathrm{net}}$ thus quantifies the residual power from available solar power after meeting the aerodynamic and transmission demands. Accordingly, during the NTO phase, when solar power is unavailable, the net power becomes negative and is given by
\begin{align}
    P_{\mathrm{net}}[n] = - P_{\mathrm{SCF}}[n] - P_{\mathrm{ave}}[n]. 
\end{align}
The evolution of the total stored energy in the batteries is then modeled as~\cite{marriott2020trajectory,bai2023adaptive}
\begin{align}
    E_{\mathrm{total}}[n+1] = E_{\mathrm{total}}[n] + \eta_b P_{\mathrm{net}}[n]\Delta_t,
\end{align}
where the battery efficiency $\eta_b$ represents either the charging efficiency $\eta_c=0.9$ when $P_{\mathrm{net}}[n] > \mu_{\mathrm{min}}$ or the inverse of the discharging efficiency, $1/\eta_d$, with $\eta_d=0.9$, when $P_{\mathrm{net}}[n]\leq \mu_{\mathrm{min}}$. Here, $\mu_{\mathrm{min}}=0$ is the minimum power threshold required to initiate battery charging, and $\Delta_t$ is the time interval between two consecutive battery states, during which the net power can be assumed approximately constant.

\section{Problem Formulation}
\label{sec_problem}
To ensure continuous flight and uninterrupted service availability, we formulate two distinct optimization problems corresponding to the DTO and NTO phases of the solar-powered HAPS. During the daytime, when solar energy is abundant, the HAPS jointly optimizes its trajectory and transmit beamforming to efficiently support both communication and SAR imaging functions while harvesting solar energy. In contrast, during the nighttime, the HAPS focuses on minimizing propulsion power consumption to maintain sustainable operation using the stored energy.

\subsection{Optimization Problem for DTO}
The DTO phase focuses on energy-aware communication and sensing optimization for the HAPS. During daylight hours, the HAPS simultaneously serves ground users and performs SAR imaging while harvesting solar energy. The objective of the DTO problem is to maximize the average achievable sum-rate by jointly optimizing the flight speed profile, transmit beamforming vectors, and sensing covariance matrix, subject to sensing, mobility, power, and energy constraints. Since the figure-eight trajectory is predetermined, the HAPS position $\mathbf{p}[n]$ is uniquely determined by the flight speed profile through the angular recursions in (\ref{trajectory_cons1}) and is therefore not treated as an independent optimization variable. The optimization problem is formulated as follows:
\begin{subequations}\label{P1}
\begin{align}
\mathcal{P}1: &\mathop{\max}\limits_{V[n],\mathbf{w}_k[n],\atop\mathbf{R}_s[n]\succeq\pmb{0},\forall n} \frac{1}{N}\sum_{n=1}^N\sum_{k=1}^K\mathcal{R}_k[n], \notag\\
\mathrm{s.t.}&\,\, \mathbf{a}^H(\mathbf{p}[n],\mathbf{t}_q)\left(\sum_{k=1}^K\mathbf{w}_k[n]\mathbf{w}_k^H[n]+\mathbf{R}_s[n]\right)\notag\\
&\qquad\quad \times\mathbf{a}(\mathbf{p}[n],\mathbf{t}_q)\geq \Gamma d^2(\mathbf{p}[n],\mathbf{t}_q), \,\,\forall q,  \tag{\ref{P1}a}\\
&\,\, \mathrm{SNR}_{\psi}[n]\geq \mathrm{SNR}_{\mathrm{min}},   \tag{\ref{P1}b} \\
&\,\, \sum_{k=1}^K\|\mathbf{w}_k[n]\|^2 + \mathrm{tr}(\mathbf{R}_s[n]) \leq P_{\mathrm{max}}, \tag{\ref{P1}c} \\
&\,\, (\ref{trajectory_cons3}),\, \sum_{n=1}^{N_1}V[n]=\frac{2\pi R_1}{\Delta_t},\,\sum_{n=N_1+1}^{N}V[n]=\frac{2\pi R_2}{\Delta_t}, \tag{\ref{P1}d} \\
&\,\, P_a -  P_{\mathrm{SCF}}[n] - P_{\mathrm{ave}}[n] \geq 0, \tag{\ref{P1}e} \\
&\,\, E_{\mathrm{total}}[N]\geq e_{\mathrm{req}}, \tag{\ref{P1}f}
\end{align}
\end{subequations}
where constraint~(\ref{P1}a) ensures that the sensing beampattern gain toward each target exceeds the prescribed threshold, thereby providing sufficient illumination for SAR imaging. Constraint~(\ref{P1}b) guarantees that the received SAR SNR satisfies the minimum imaging requirement, whereas constraint~(\ref{P1}c) limits the total communication and sensing transmit power. Constraint~(\ref{P1}d) specifies the feasible flight-speed range and ensures that the HAPS completes one full revolution on each circle of the prescribed figure-eight trajectory. Constraint~(\ref{P1}e) enforces the instantaneous power balance during the DTO phase by requiring the harvested solar power to simultaneously support propulsion and communication/sensing transmissions, thereby preventing battery discharge. Finally, constraint~(\ref{P1}f) guarantees that the battery energy remaining at the end of the DTO phase is no smaller than the minimum energy required for the subsequent NTO phase, denoted by $e_{\mathrm{req}}$, thereby coupling Problems~$\mathcal{P}1$ and~$\mathcal{P}2$. The solution to Problem~$\mathcal{P}1$ is presented in Section~\ref{sec_method_DTO}.

\begin{remark}
The available solar power $P_a$ is mainly determined by the geographic latitude, seasonal variation, and the effective solar-panel area of the HAPS. As illustrated in Fig.~\ref{E_days}, regions located at lower latitudes can harvest sufficient solar energy throughout the year, whereas at higher latitudes, efficient battery charging and energy storage can only be achieved during periods of high irradiance. During low-irradiance seasons, constraint~(\ref{P1}e) may not be satisfied, as the harvested solar energy can fall below the total power required for SCF and payload operation. In addition, since the PV modules are typically mounted on the upper surfaces of the wings~\cite{sornek2025status}, the total effective collection area $A_p$ is inherently limited by the available wing surface area $S$. In this work, we simply assume $A_p=S=143$ m$^{2}$.

As illustrated in Fig.~\ref{It_hours}, solar irradiance remains approximately stable during the central daylight hours. Since the DTO optimization considers a single figure-eight flight cycle, variations in harvested solar power over this short period can be reasonably neglected under the considered operating conditions. For longer operating periods or significant fluctuations in solar irradiance, dynamic energy arrivals may affect the instantaneous power balance and battery energy evolution, thereby influencing the feasibility of the energy constraints.
\end{remark}

\subsection{Optimization Problem for NTO}
After completing the DTO phase, the battery energy available at the beginning of the NTO phase is denoted by $e_{\mathrm{start}}=E_{\mathrm{total}}[N]$. Since solar energy is unavailable during the NTO phase, the HAPS relies exclusively on the stored battery energy to support propulsion, communication, and SAR imaging. To maximize the nighttime endurance, the total propulsion power consumption is minimized subject to the flight and battery-energy constraints. The resulting optimization problem is formulated as
\begin{subequations}\label{P2}
\begin{align}
    \mathcal{P}2:&\mathop{\min}_{H,V[n],\forall n} \sum_{n=1}^{N}P_{\mathrm{SCF}}(V[n],H), \notag\\
    &\quad\,\,\, \mathrm{s.t.}\,\, H_{\rm{min}}\leq H\leq H_{\rm{max}}, \tag{\ref{P2}a}\\
    &\qquad\quad\, V_{\rm{min}}\leq V[n]\leq V_{\rm{max}},  \tag{\ref{P2}b}\\
    &\qquad\quad\, \sum_{n=1}^N\frac{(P_{\mathrm{SCF}}[n] + P_{\mathrm{ave}}[n])\Delta_t}{\eta_d} \leq e_{\mathrm{start}},  \tag{\ref{P2}c}
\end{align}
\end{subequations}
where constraints~(\ref{P2}a) and~(\ref{P2}b) specify the allowable flight altitude and speed ranges, respectively. Constraint~(\ref{P2}c) ensures that the total battery energy consumed during the NTO phase does not exceed the available battery energy at its beginning. Since $e_{\mathrm{start}}=E_{\mathrm{total}}[N]$ and constraint~(\ref{P1}f) guarantees $E_{\mathrm{total}}[N]\geq e_{\mathrm{req}}$, the battery energy available at the beginning of the NTO phase always satisfies its minimum energy requirement. Consequently, Problems~$\mathcal{P}1$ and~$\mathcal{P}2$ are coupled through constraints~(\ref{P1}f) and~(\ref{P2}c), which govern the battery energy transferred from the DTO phase to the NTO phase. The solution to Problem~$\mathcal{P}2$ is presented in Section~\ref{sec_method_NTO}.


\par

\section{Proposed Algorithm for DTO}
\label{sec_method_DTO}
This section develops an alternating optimization (AO) algorithm based on the successive convex approximation (SCA) framework to solve Problem~$\mathcal{P}1$. The proposed algorithm alternates between two optimization subproblems. Specifically, for a given HAPS trajectory, the communication beamforming vectors $\mathbf{w}_k[n]$ and the SAR imaging covariance matrices $\mathbf{R}_s[n]$ are first optimized. Subsequently, the flight speed profile is updated with the optimized beamforming and sensing variables while satisfying the predefined figure-eight trajectory constraints. Since the HAPS position is uniquely determined by the flight speed profile, the trajectory is implicitly updated during this step. The above procedure is repeated until the objective value converges.

\subsection{Transmit Beamforming Optimization}
Given the HAPS trajectory $\mathbf{p}[n]$, which is uniquely determined by the flight speed profile $V[n]$, the objective is to jointly optimize the communication beamforming vectors $\mathbf{w}_k[n]$ and the SAR imaging covariance matrices $\mathbf{R}_s[n]$. The corresponding optimization problem is formulated as
\begin{align}
\mathcal{P}3: &\mathop{\max}\limits_{\mathbf{w}_k[n],\forall k,\atop\mathbf{R}_s[n]\succeq\pmb{0},\forall n}\,\frac{1}{N}\sum_{n=1}^N\sum_{k=1}^K \mathcal{R}_k[n], \notag\\
&\quad\,\,\,\,\mathrm{s.t.}\quad (\ref{P1}a), (\ref{P1}b), (\ref{P1}c), (\ref{P1}e)\,\mathrm{and}\,(\ref{P1}f). \notag
\end{align}
For fixed $\mathbf{p}[n]$ and $V[n]$, constraints (\ref{P1}a)-(\ref{P1}c) and (\ref{P1}e) are separable across time slots, whereas the terminal battery energy constraint (\ref{P1}f) couples the transmit powers across time slots. As (\ref{P1}e) ensures $P_{\mathrm{net}}[n]\geq0$, only the charging branch of the battery energy model is active during the DTO phase. Consequently, (\ref{P1}f) is equivalent to the following aggregate power constraint:
\begin{align}
\sum_{n=1}^{N}P_{\mathrm{ave}}[n]\leq\bar E(V),
\label{sum_power}
\end{align}
where $E_{\mathrm{total}}[1]$ denotes the initial battery energy of the DTO phase, and $\bar E(V)\triangleq
\sum_{n=1}^{N}\big(P_a-P_{\mathrm{SCF}}[n]\big)
-\frac{e_{\mathrm{req}}-E_{\mathrm{total}}[1]}
{\eta_c\Delta_t}$ is constant for a fixed speed profile. To handle this coupling, we introduce a common Lagrange multiplier $\lambda\geq0$ associated with (\ref{sum_power}),  with the normalization factor $1/N$ in the objective of $\mathcal{P}3$ absorbed into the multiplier. For a given $\lambda$, the resulting Lagrangian maximization problem can be decomposed into $N$ independent per-slot subproblems as follows:
\begin{align}
\mathcal{P}4.n:
&\mathop{\max}\limits_{\mathbf{w}_k[n],\forall k,
\atop\mathbf{R}_s[n]\succeq\pmb{0}}\,
\sum_{k=1}^K \mathcal{R}_k[n]
-\lambda P_{\mathrm{ave}}[n], \notag\\
&\quad\mathrm{s.t.}\quad
(\ref{P1}a),(\ref{P1}b),(\ref{P1}c)
\,\mathrm{and}\,(\ref{P1}e). \notag
\end{align}
The multiplier $\lambda$ serves as a common power price that coordinates the transmit power allocation across time slots. Since $\mathcal{P}4.n$ is non-convex, the multiplier is determined at each SCA iteration based on the convex SDR subproblems derived below.

For problem $\mathcal{P}4.n$, we introduce the auxiliary matrix variable 
$\mathbf{W}_k[n] = \mathbf{w}_k[n]\mathbf{w}_k[n]^{H}$, 
which satisfies $\mathbf{W}_k[n] \succeq \pmb{0}$ and $\mathrm{rank}(\mathbf{W}_k[n])\leq 1$. Accordingly, problem {$\mathcal{P}4.n$} can be equivalently reformulated as
\begin{subequations}\label{P5}
\begin{align}
\mathcal{P}5.n: &\mathop{\max}\limits_{\mathbf{W}_k[n]\succeq\pmb{0},\forall k,\atop\mathbf{R}_s[n]\succeq\pmb{0}}\,\sum_{k=1}^K \hat{\mathcal{R}}_k(\{\mathbf{W}_k[n]\},\mathbf{R}_s[n])-\lambda P_{\mathrm{ave}}[n],\notag\\
&\mathrm{s.t.}\,\mathbf{a}^H(\mathbf{p}[n],\mathbf{t}_q)\left(\sum_{k=1}^K\mathbf{W}_k[n]+\mathbf{R}_s[n]\right)\mathbf{a}(\mathbf{p}[n],\mathbf{t}_q)\notag\\
&\qquad\qquad\qquad\qquad\qquad\geq \Gamma d^2(\mathbf{p}[n],\mathbf{t}_q), \,\,\forall q,  \tag{\ref{P5}a}\\
&\sum_{k=1}^K\mathrm{tr}(\mathbf{W}_k[n]) + \mathrm{tr}(\mathbf{R}_s[n])\geq \frac{\mathrm{SNR}_{\mathrm{min}}V[n]}{C_{\mathrm{SAR}}\sin(\psi)^2},  \tag{\ref{P5}b} \\
&\sum_{k=1}^K\mathrm{tr}(\mathbf{W}_k[n]) + \mathrm{tr}(\mathbf{R}_s[n]) \leq P_{\mathrm{max}}, \tag{\ref{P5}c} \\
&\mathrm{rank}(\mathbf{W}_k[n]) \leq 1,\tag{\ref{P5}d} \\
&\sum_{k=1}^K\mathrm{tr}(\mathbf{W}_k[n]) + \mathrm{tr}(\mathbf{R}_s[n])\leq P_a-P_{\mathrm{SCF}}[n], \tag{\ref{P5}e}
\end{align}
\end{subequations}
where $\hat{\mathcal{R}}_k(\{\mathbf{W}_k[n]\},\mathbf{R}_s[n])$ is given by \begin{align}
&\hat{\mathcal{R}}_k(\{\mathbf{W}_k[n]\},\mathbf{R}_s[n]) \notag\\
    &= \log_2\Big(\sum\limits_{p=1}^K\mathrm{tr}\left(\mathbf{G}_k\mathbf{W}_p[n]\right)+\mathrm{tr}\left(\mathbf{G}_k\mathbf{R}_s[n]\right) + \sigma_k^2 \Big) \notag\\
    &\,\,\, -\log_2\Big(\sum\limits_{p=1\atop p\neq k}^K\mathrm{tr}\left(\mathbf{G}_k\mathbf{W}_p[n]\right)+\mathrm{tr}\left(\mathbf{G}_k\mathbf{R}_s[n]\right) + \sigma_k^2 \Big). \notag
\end{align}   
with
\begin{align}
    \mathbf{G}_k =& \mathbb{E}\{\mathbf{g}(\mathbf{p}[n],\mathbf{u}_k)\mathbf{g}^H(\mathbf{p}[n],\mathbf{u}_k)\}\notag\\
    =&\frac{1}{\ell (K_u+1)}(K_u\mathbf{a}_k(\mathbf{p}[n],\mathbf{u}_k)\mathbf{a}_k^H(\mathbf{p}[n],\mathbf{u}_k)+\mathbf{I}_M). 
\end{align}

Because the objective function of $\mathcal{P}5.n$ is non-concave and the rank constraints in (\ref{P5}d) are non-convex, problem $\mathcal{P}5.n$ remains a challenging non-convex optimization problem. To tackle this issue, we adopt the SCA method~\cite{dinh2010local,zeng2016throughput,zeng2017energy}, which iteratively replaces the non-concave objective with a concave surrogate function, i.e., a locally tight and tractable approximation that preserves the original function’s first-order behavior. Specifically, at each iteration $o \geq 1$, we have $\hat{\mathcal{R}}_k(\{\mathbf{W}_k[n]\},\mathbf{R}_s[n]) \geq \bar{\mathcal{R}}_k^{(o)}(\{\mathbf{W}_k[n]\},\mathbf{R}_s[n])$, where $\bar{\mathcal{R}}_k^{(o)}({\mathbf{W}_k[n]},\mathbf{R}_s[n])$ denotes the concave surrogate of $\hat{\mathcal{R}}_k$ at iteration $o$, which can be expressed as in (\ref{R_k_iteration}) with 
\begin{figure*}
\begin{align}\label{R_k_iteration}
    \bar{\mathcal{R}}_k^{(o)}(\{\mathbf{W}_k[n]\},\mathbf{R}_s[n]) =& \log_2\Big(\sum\limits_{p=1}^K\mathrm{tr}\left(\mathbf{G}_k\mathbf{W}_p[n]\right)+\mathrm{tr}\left(\mathbf{G}_k\mathbf{R}_s[n]\right) + \sigma_k^2\Big) -\bar{a}_k^{(o)}[n] \notag\\
    &-\sum_{p=1,p\neq k}^K\mathrm{tr}\Big(\bar{\mathbf{B}}_k^{(o)}[n](\mathbf{W}_p[n]-\mathbf{W}_p^{(o)}[n]\big)\Big)-\mathrm{tr}\left(\bar{\mathbf{B}}_k^{(o)}[n](\mathbf{R}_s[n]-\mathbf{R}_s^{(o)}[n])\right).
\end{align}   
\hrulefill
\end{figure*}

\begin{align}
    \bar{a}_k^{(o)}[n] = \log_2\Big(\sum\limits_{p=1\atop p\neq k}^K\mathrm{tr}\Big(\mathbf{G}_k\mathbf{W}_p^{(o)}[n]\Big)+\mathrm{tr}\left(\mathbf{G}_k\mathbf{R}_s^{(o)}[n]\right) + \sigma_k^2 \Big). \notag
\end{align}
\begin{align}
    \bar{\mathbf{B}}_k^{(o)}[n] = \frac{\log_2(e)\mathbf{G}_k}{\sum\limits_{p=1\atop p\neq k}^K\mathrm{tr}\left(\mathbf{G}_k\mathbf{W}_p^{(o)}[n]\right)+\mathrm{tr}\Big(\mathbf{G}_k\mathbf{R}_s^{(o)}[n]\Big)+\sigma_k^2}. \notag
\end{align}
Furthermore, following the semidefinite relaxation (SDR) technique in~\cite{luo2010semidefinite,lyu2022joint}, the rank constraints in~(\ref{P5}d) are relaxed, transforming problem $\mathcal{P}5.n$ into a convex semidefinite program (SDP) for a given $\lambda$, denoted as (SDR5.n). At each SCA iteration, the $N$ convex subproblems can be solved in parallel using CVX~\cite{grant2016cvx}, with a common multiplier $\lambda$ determined as follows.

\textit{Determination of $\lambda$:}
At each SCA iteration, with the linearization points fixed, let $P_{\mathrm{ave}}^{\star}[n;\lambda]$ denote the transmit power corresponding to an optimal solution of (SDR5.n) 
for a given $\lambda$. The aggregate transmit power is defined as $P_\Sigma(\lambda)\triangleq
\sum_{n=1}^{N}P_{\mathrm{ave}}^{\star}[n;\lambda]$. Since $P_\Sigma(\lambda)$ is non-increasing in $\lambda$, we set $\lambda=0$ if $P_\Sigma(0)\leq\bar E(V)$. Otherwise, a positive multiplier is needed to reduce the aggregate transmit power. Let $P_{\min}[n]$ denote the minimum transmit power over the feasible set of (SDR5.n), and define $P_{\min}\triangleq\sum_{n=1}^{N}P_{\min}[n]$. 
Feasible initialization and updates ensure 
$P_{\min}\leq\bar E(V)$. If $P_{\min}=\bar E(V)$, 
each time slot must operate at its minimum transmit power. In this case, the subproblems are solved with $\lambda=0$ and the additional constraints $P_{\mathrm{ave}}[n]\leq P_{\min}[n]$, $\forall n$. If $P_{\min}<\bar E(V)$, a finite search interval $[\lambda_{\mathrm l},\lambda_{\mathrm u}]$ satisfying $P_\Sigma(\lambda_{\mathrm l})>
\bar E(V)\geq P_\Sigma(\lambda_{\mathrm u})$ exists whenever $P_\Sigma(0)>\bar E(V)$. 
Bisection is then performed until $\lambda_{\mathrm u}-\lambda_{\mathrm l}\leq\epsilon_\lambda$, where $\epsilon_\lambda>0$ is a prescribed tolerance. After bisection, the aggregate power constraint 
may not hold with equality due to finite search 
accuracy or non-unique optimal solutions. To address this issue, we construct a feasible solution as $\mathbf X[n]=\theta\mathbf X_{\mathrm l}[n]
+(1-\theta)\mathbf X_{\mathrm u}[n]$ with $\theta=\frac{\bar E(V)-P_\Sigma(\lambda_{\mathrm u})}{P_\Sigma(\lambda_{\mathrm l})-P_\Sigma(\lambda_{\mathrm u})}\in[0,1]$. Here, $\mathbf X[n]$ collectively denotes 
$\{\mathbf W_k[n]\}_{k=1}^{K}$ and $\mathbf R_s[n]$, 
and $\mathbf X_{\mathrm l}[n]$ and $\mathbf X_{\mathrm u}[n]$ are the solutions at the final bisection endpoints. The convexity of the relaxed feasible sets ensures that the constructed solution satisfies all per-slot constraints, while the linearity of $P_{\mathrm{ave}}[n]$ ensures that (\ref{sum_power}) holds with equality.

To ensure monotonicity, the resulting solution is accepted only if its surrogate sum-rate is no lower than that of the current iterate, evaluated using the same surrogate function; otherwise, the current iterate is retained. Repeating this procedure yields a sequence of solutions $\mathbf W_k^{(o)}[n]$ and $\mathbf R_s^{(o)}[n]$ satisfying the terminal battery energy constraint (\ref{P1}f) without preallocating energy budgets across time slots.

\par

\subsection{Trajectory Optimization}
Given the communication beamforming vectors $\mathbf{w}_k[n]$ and the SAR imaging covariance matrix $\mathbf{R}_s[n]$, our objective is to optimize the HAPS flight speed profile $V[n]$ along the predefined figure-eight geometry, which in turn determines the HAPS trajectory $\mathbf{p}[n]$ through the angular recursion in (\ref{trajectory_cons1}). This leads to the following trajectory optimization problem, denoted as $\mathcal{P}6$, formulated as
\begin{subequations}\label{P6}
\begin{align}
\mathcal{P}6:&\mathop{\max}\limits_{V[n],\forall n}\,\frac{1}{N}\sum_{n=1}^N\sum_{k=1}^K \mathcal{R}_k[n], \notag\\
&\mathrm{s.t.}\,\, \mathbf{a}^H(\mathbf{p}[n],\mathbf{t}_q)\mathbf{D}[n]\mathbf{a}(\mathbf{p}[n],\mathbf{t}_q)\notag\\
&\qquad\qquad\qquad\,\, \geq \Gamma(H^2+\|\mathbf{p}[n]-\mathbf{t}_q\|^2), \,\,\forall q,  \\
&V[n]\leq \frac{P_{\mathrm{ave}}[n]C_{\mathrm{SAR}}\sin^2(\psi)}{\mathrm{SNR}_{\mathrm{min}}},\,(\ref{P1}d)-(\ref{P1}f),  
\end{align}
\end{subequations}
where $\mathbf{D}[n]=\sum_{k=1}^K\mathbf{w}_k[n]\mathbf{w}_k^H[n]+\mathbf{R}_s[n]$, and $\mathbf{p}[n]$ is understood throughout as the function of $V[1],\ldots,V[n-1]$ given by the recursion in (\ref{trajectory_cons1}). Since the propulsion power $P_{\mathrm{SCF}}[n]$ is a convex function of the flight speed $V[n]$~\cite{javed2023interdisciplinary}, constraints~(\ref{P1}d)-(\ref{P1}f) are all convex with respect to $V[n]$. However, the overall problem $\mathcal{P}6$ remains non-convex due to the non-concave objective function and the non-convex constraint (\ref{P6}a). To overcome this challenge, we develop a trust-region-based SCA algorithm to iteratively approximate and solve $\mathcal{P}6$ with non-decreasing objective values.

Firstly, let $\mathbf{W}_k[n]=\mathbf{w}_k[n]\mathbf{w}_k^H[n]$ and $\mathbf{A}_k=\mathbf{a}(\mathbf{p}[n],\mathbf{u}_k)\mathbf{a}^H(\mathbf{p}[n],\mathbf{u}_k)$, we can re-express the $\mathcal{R}_k[n]$ in $\mathcal{P}6$ as 
\begin{align}\label{P6_objective}
    \hat{\mathcal{R}}_k[n] =& \log_2(\eta_k[n]) -\log_2(\mu_k[n]),
\end{align}
where
\begin{align}
    &\eta_k[n] = \sum_{p=1}^Kf(\mathbf{W}_p[n],d(\mathbf{p}[n],\mathbf{u}_k)) \notag \\
    &\,\,\,+ g(\mathbf{R}_s[n],d(\mathbf{p}[n],\mathbf{u}_k)) + \frac{\sigma_k^2(K_u+1)}{\rho_0 K_u}d^2(\mathbf{p}[n],\mathbf{u}_k), \\
    &\mu_k[n] = \sum_{p=1, p\neq 
 k}^Kf(\mathbf{W}_p[n],d(\mathbf{p}[n],\mathbf{u}_k)) \notag \\
    &\,\,\,+ g(\mathbf{R}_s[n],d(\mathbf{p}[n],\mathbf{u}_k)) + \frac{\sigma_k^2(K_u+1)}{\rho_0 K_u}d^2(\mathbf{p}[n],\mathbf{u}_k), 
\end{align}
with 
\begin{align}
    &f\left(\mathbf{W}_k[n],d(\mathbf{p}[n],\mathbf{u}_k)\right) = \mathrm{tr}((\mathbf{A}_k+K_u^{-1}\mathbf{I}_M)\mathbf{W}_k[n]) \notag\\
    &= \sum_{i=1}^M\sum_{j=1}^M[\mathbf{W}_k[n]]_{i,j}e^{\frac{j\pi(j-i)H}{d(\mathbf{p}[n],\mathbf{u}_k)}} + K_u^{-1}\mathrm{tr}(\mathbf{W}_k[n]) \notag\\
    &= \frac{K_u+1}{K_u}\mathrm{tr}(\mathbf{W}_k[n]) + 2\sum_{i=1}^M\sum_{j=i+1}^M|[\mathbf{W}_k[n]]_{i,j}| \notag \\
    &\quad\times\cos\left(\theta_{i,j}^{W_{k}}[n] + \frac{\pi(j-i)H}{d(\mathbf{p}[n],\mathbf{u}_k)}\right),
\end{align}
\begin{align}
    &g\left(\mathbf{R}_s[n],d(\mathbf{p}[n],\mathbf{u}_k)\right) = \mathrm{tr}((\mathbf{A}_k+K_u^{-1}\mathbf{I}_M)\mathbf{R}_s[n])\notag\\
    &= \sum_{i=1}^M\sum_{j=1}^M[\mathbf{R}_s[n]]_{i,j}e^{\frac{j\pi(j-i)H}{d(\mathbf{p}[n],\mathbf{u}_k)}} + K_u^{-1}\mathrm{tr}(\mathbf{R}_s[n]) \notag\\
    &= \frac{K_u+1}{K_u}\mathrm{tr}(\mathbf{R}_s[n]) + 2\sum_{i=1}^M\sum_{j=i+1}^M|[\mathbf{R}_s[n]]_{i,j}| \notag \\
    &\quad\times\cos\left(\theta_{i,j}^{R_s}[n] + \frac{\pi (j-i)H}{d(\mathbf{p}[n],\mathbf{u}_k)}\right),
\end{align}
with $\theta_{i,j}^{W_k}[n]$ and $\theta_{i,j}^{R_s}[n]$ being the phases, and $|[\mathbf{W}_k[n]]_{i,j}|$ and $|[\mathbf{R}_s[n]]_{i,j}|$ being the magnitudes of the $(i,j)$th entries of $\mathbf{W}_k[n]$ and $\mathbf{R}_s[n]$, respectively. Similarly, the non-convex constraints in (\ref{P6}a) can be rewritten as 
\begin{align}\label{P6_constraint}
    &\mathrm{tr}(\mathbf{D}[n]) + 2\sum_{i=1}^M\sum_{j=i+1}^M|[\mathbf{D}[n]]_{i,j}| \notag\\
    &\quad\times\cos\left(\theta_{i,j}^{D}[n] + \frac{\pi(j-i)H}{d(\mathbf{p}[n],\mathbf{t}_q)}\right) \geq \Gamma d^2(\mathbf{p}[n],\mathbf{t}_q), 
\end{align}
where $\theta_{i,j}^{D}[n]$ and $|[\mathbf{D}[n]]_{i,j}|$ are the phase and magnitude of the $(i,j)$th entries of $\mathbf{D}[n]$, respectively.

Next, we employ a first-order Taylor expansion to linearize the non-concave objective function with respect to $\mathbf{p}[n]$, i.e.,
\begin{align}\label{Taylor_objective}
    \hat{\mathcal{R}}_k[n] \approx \bar{\mathcal{R}}_k[n] =& r_k^{(l)}[n] + \mathbf{v}_k^{(l)H}[n](\mathbf{p}[n]-\mathbf{p}^{(l)}[n]), 
\end{align}
where $r_k^{(l)}[n] = \log_2(\eta_k^{(l)}[n]) -\log_2(\mu_k^{(l)}[n])$, and $\mathbf{v}_k^{(l)}[n]$ is expressed as (\ref{v_k}) with 
\begin{figure*}
\begin{align}
    \mathbf{v}_k^{(l)}[n] =& \frac{\log_2(e)}{\eta_k^{(l)}[n]}\Big(\sum_{p=1}^K\pmb{\gamma}_p^{(l)}[n]+\pmb{\omega}_s^{(l)}[n]+\frac{2\sigma_k^2(K_u+1)(\mathbf{p}^{(l)}[n]-\mathbf{u}_k)}{\rho_0K_u}\Big) \notag\\
    &- \frac{\log_2(e)}{\mu_k^{(l)}[n]}\Big(\sum_{p=1,p\neq k}^K\pmb{\gamma}_p^{(l)}[n]+\pmb{\omega}_s^{(l)}[n]+\frac{2\sigma_k^2(K_u+1)(\mathbf{p}^{(l)}[n]-\mathbf{u}_k)}{\rho_0 K_u}\Big). \label{v_k} 
\end{align}
\hrulefill
\end{figure*}
 
\begin{align}
\pmb{\gamma}_{p}^{(l)}[n] =& \sum_{i=1}^M\sum_{j=i+1}^M2\pi|[\mathbf{W}_{p}[n]]_{i,j}|\sin\left(\theta_{i,j}^{W_{p}}[n] \right. \notag\\
    &\left.+\frac{\pi (j-i)H}{d(\mathbf{p}^{(l)}[n],\mathbf{u}_k)}\right)\frac{(j-i)H(\mathbf{p}^{(l)}[n]-\mathbf{u}_k)}{d^3(\mathbf{p}^{(l)}[n],\mathbf{u}_k)}, \\
    \pmb{\omega}_s^{(l)}[n] =& \sum_{i=1}^M\sum_{j=i+1}^M2\pi|[\mathbf{R}_s[n]]_{i,j}|\sin\left(\theta_{i,j}^{R_s}[n]\right. \notag\\
    &\left.+\frac{\pi (j-i)H}{d(\mathbf{p}^{(l)}[n],\mathbf{u}_k)}\right)\frac{(j-i)H(\mathbf{p}^{(l)}[n]-\mathbf{u}_k)}{d^3(\mathbf{p}^{(l)}[n],\mathbf{u}_k)}, 
\end{align}
Similarly, applying the first-order Taylor expansion to the non-convex constraints in $(\ref{P6}a)$ yields
\begin{align}\label{Taylor_constraint}
    &\Xi_q^{(l)}[n] + \pmb{\nu}_q^{(l)H}[n](\mathbf{p}[n]-\mathbf{p}^{(l)}[n]) \geq \Gamma(H^2+\|\mathbf{p}[n]-\mathbf{t}_q\|^2), 
\end{align}
where 
\begin{align}
    \Xi_q^{(l)}[n] =& \mathrm{tr}(\mathbf{D}[n]) + 2\sum_{i=1}^M\sum_{j=i+1}^M|[\mathbf{D}[n]]_{i,j}| \notag\\
    &\times\cos\left(\theta_{i,j}^{D}[n] + \frac{\pi (j-i)H}{d(\mathbf{p}^{(l)}[n],\mathbf{t}_q)}\right), \\
    \pmb{\nu}_q^{(l)}[n] =& \sum_{i=1}^M\sum_{j=i+1}^M2\pi|[\mathbf{D}[n]]_{i,j}|\sin\left(\theta_{i,j}^{D}[n]  \right. \notag\\
    &\left.+ \frac{\pi(j-i)H}{d(\mathbf{p}^{(l)}[n],\mathbf{t}_q)}\right)\frac{(j-i)H(\mathbf{p}^{(l)}[n]-\mathbf{t}_q)}{d^3(\mathbf{p}^{(l)}[n],\mathbf{t}_q)}.
\end{align}
At this stage, both the non-concave objective function in (\ref{P6_objective}) and the non-convex constraint in (\ref{P6_constraint}) are approximated by their respective linearized forms in (\ref{Taylor_objective}) and (\ref{Taylor_constraint}), both of which remain affine in the position increment $\mathbf{p}[n]-\mathbf{p}^{(l)}[n]$.

Since the HAPS position $\mathbf{p}[n]$ is uniquely determined by the flight speed profile through the angular recursions in~(\ref{trajectory_cons1}), it is no longer treated as an independent optimization variable. Instead, the position increment is expressed in terms of the flight-speed increment. Specifically, define the accumulated flight-speed increment as
\begin{align}
&\Delta s^{(l)}[n] \notag\\
&\triangleq\left\{
\begin{array}{ll}
   \Delta_t\sum\limits_{m=1}^{n-1}\Big(V[m]-V^{(l)}[m]\Big),  &  n=1,\ldots,N_1,\\
   \Delta_t\sum\limits_{m=N_1+1}^{n-1}\Big(V[m]-V^{(l)}[m]\Big),  &  n=N_1+1,\ldots,N,
\end{array}\right.  \notag
\end{align}
where the two summation intervals correspond to the two circular segments of the predefined figure-eight trajectory. 
Let $\boldsymbol{\tau}^{(l)}[n]$ denote the unit tangent vector of the corresponding circular arc evaluated at the current iterate $\mathbf{p}^{(l)}[n]$, i.e.,
\begin{align}
\boldsymbol{\tau}^{(l)}[n]=
\begin{cases}
[\cos\varphi_1^{(l)}[n],-\sin\varphi_1^{(l)}[n]]^T,
& n=1,\ldots,N_1,\\
[\cos\varphi_2^{(l)}[n],\sin\varphi_2^{(l)}[n]]^T,
& n=N_1+1,\ldots,N.
\end{cases} \notag
\end{align}
Then, by applying the first-order approximation of the trajectory with respect to the flight speed, the position increment is given by
\begin{align}\label{p_to_V}
\mathbf{p}[n]-\mathbf{p}^{(l)}[n]
\approx
\boldsymbol{\tau}^{(l)}[n]\Delta s^{(l)}[n].
\end{align}
Substituting~(\ref{p_to_V}) into~(\ref{Taylor_objective}) and~(\ref{Taylor_constraint}) yields the following affine approximations:
\begin{align}
&\bar{\mathcal{R}}_k[n] = r_k^{(l)}[n] + \mathbf{v}_k^{(l)H}[n]\boldsymbol{\tau}^{(l)}[n]\Delta s^{(l)}[n], \label{Taylor_objective_V} \\
&\Xi_q^{(l)}[n] + \pmb{\nu}_q^{(l)H}[n]\boldsymbol{\tau}^{(l)}[n]\Delta s^{(l)}[n]
\notag\\
&\qquad\ge
\Gamma(H^2+\|\mathbf{p}^{(l)}[n] + \boldsymbol{\tau}^{(l)}[n]\Delta s^{(l)}[n]-\mathbf{t}_q\|^2).\label{Taylor_constraint_V}
\end{align}

To maintain the accuracy of the first-order approximations, the following trust-region constraints are imposed on the flight speed profile:
\begin{align}\label{trust region}
    |V^{(l+1)}[n]-V^{(l)}[n]| \leq \Delta_V^{(l)}, \forall n
\end{align}
where $\Delta_V^{(l)}$ denotes the trust-region radius at the $l$th iteration.

Finally, by combining (\ref{Taylor_objective_V}), (\ref{Taylor_constraint_V}), and (\ref{trust region}), we obtain the convex approximation of problem $\mathcal{P}6$ in the $l$th iteration, denoted as problem $\mathcal{P}7.l$, i.e.,
\begin{align}
    &\mathcal{P}7.l: \max_{V[n],\forall n} \frac{1}{N}\sum_{n=1}^N\sum_{k=1}^K\bar{\mathcal{R}}_k[n],\notag\\
    &\qquad\quad\,\,\,\mathrm{s.t.}\,\,\, (\ref{Taylor_constraint_V}), (\ref{trust region})\,\mathrm{and}\,\, (\ref{P6}b), \notag
\end{align}
which can be efficiently solved to optimality using CVX. Note that a sufficiently small trust 
region radius $\Delta_{V}^{(l)}$ helps control the 
local approximation error~\cite{conn2000trust}. 
In practical implementation, however, if solving 
$\mathcal{P}7.l$ in the $l$th iteration yields a solution that violates any constraint of $\mathcal{P}6$ or results in a lower 
objective value for $\mathcal{P}6$ than that of the 
current iterate, we reduce the trust region radius 
to $\Delta_{V}^{(l)}/2$ and solve $\mathcal{P}7.l$ again. 
The iteration process terminates when 
$\Delta_{V}^{(l)}$ falls below a predefined threshold 
$\zeta$, with the last accepted iterate 
retained.

\begin{algorithm}[t]
\caption{DTO - Joint Beamforming and Trajectory Design}
\label{alg_DTO}
\begin{algorithmic}[1]
    \REQUIRE $H$, $T$, $N$, $\Delta_t$, $\alpha$, $\beta$, $\mathbf{u}_k$, $\forall k$, $\mathbf{t}_q$, $\forall q$, $\mathrm{SNR}_{\mathrm{min}}$, $\Gamma$, $P_{\mathrm{max}}$, $P_a$, $V_{\mathrm{min}}$, and $V_{\mathrm{max}}$.
    \ENSURE $\mathbf{w}_k^{\star}[n]$, $\mathbf{R}_s^{\star}[n]$, $V^{\star}[n]$, and $\mathbf{p}^{\star}[n]$, $\forall k,n$.
    \STATE Feasibly initialize $\mathbf{W}_k^{(0)}[n]$, $\mathbf{R}_s^{(0)}[n]$, and $V^{(0)}[n]$, $\forall k,n$, and compute the corresponding initial trajectory $\mathbf{p}^{(0)}[n]$ via the angular recursion in (\ref{trajectory_cons1}). Let $o=0$.
    \WHILE{no convergence}
        \STATE Update $\mathbf W_k^{(o+1)}[n]$ and $\mathbf R_s^{(o+1)}[n]$ using the SCA-SDR subproblems (SDR5.n) and the common-multiplier procedure for fixed $V^{(o)}[n]$.
        \STATE Let $l=0$, $V^{(l)}[n]=V^{(o)}[n]$, and $\mathbf{p}^{(l)}[n]=\mathbf{p}^{(o)}[n]$.
        \WHILE{no convergence}
            \STATE Obtain a candidate speed profile $\widetilde{V}^{(l+1)}[n]$ by solving $\mathcal{P}7.l$ using $\mathbf{W}_k^{(o+1)}[n]$ and $\mathbf{R}_s^{(o+1)}[n]$, $\forall k,n$.
            \STATE Compute the corresponding candidate trajectory $\widetilde{\mathbf{p}}^{(l+1)}[n]$ from $\widetilde{V}^{(l+1)}[n]$ via (\ref{trajectory_cons1}).
            \IF{the candidate satisfies the original constraints and the objective of $\mathcal{P}6$ does not decrease}
                \STATE Set $V^{(l+1)}[n]=\widetilde{V}^{(l+1)}[n]$ and $\mathbf{p}^{(l+1)}[n]=\widetilde{\mathbf{p}}^{(l+1)}[n]$, and update $l=l+1$.
            \ELSE
                \STATE Execute $\Delta_V^{(l)}=\Delta_V^{(l)}/2$, and resolve $\mathcal{P}7.l$ using $V^{(l)}[n]$.
            \ENDIF
        \ENDWHILE
        \STATE Set $V^{(o+1)}[n]=V^{(l)}[n]$ and $\mathbf{p}^{(o+1)}[n]=\mathbf{p}^{(l)}[n]$, and update $o=o+1$.
    \ENDWHILE
    \STATE Recover $\mathbf{w}_k[n]$ and the corresponding sensing covariance matrices from the relaxed solution, preserving the aggregate transmit covariance.
    \STATE Return $\mathbf{w}_k^{\star}[n]=\mathbf{w}_k[n]$, $\mathbf{R}_s^{\star}[n]=\mathbf{R}_s[n]$, $V^{\star}[n]=V[n]$, and $\mathbf{p}^{\star}[n]=\mathbf{p}[n]$, $\forall k,n$.
\end{algorithmic}
\end{algorithm}

\subsection{Convergence and Complexity Analysis} 
The overall SCA-based AO framework for solving problem 
$\mathcal{P}1$ is summarized in Algorithm~\ref{alg_DTO}. 
The algorithm alternately performs transmit beamforming 
optimization (Section~\ref{sec_method_DTO}-A) and HAPS 
trajectory optimization (Section~\ref{sec_method_DTO}-B) 
in an iterative manner. In each iteration, the transmit 
beamforming subproblem is solved using the 
SCA-and-SDR-based approach, with the 
aggregate power constraint handled through a common 
multiplier. Given the optimized covariance matrices, 
the HAPS trajectory is subsequently refined by solving 
the trust-region-based SCA problem $\mathcal{P}7.l$. 
These alternating updates preserve 
feasibility and yield a non-decreasing sequence of 
objective values for the SDR-relaxed problem. Since 
the objective value is bounded above, the resulting 
objective sequence converges.

In terms of computational complexity, the beamforming 
subproblem $\mathcal{P}5.n$ involves $(K+1)$ positive 
semidefinite matrices of size $M\times M$. Using the interior-point complexity 
approximation, solving the $N$ relaxed subproblems 
for a given $\lambda$ incurs a computational cost 
on the order of $\mathcal{O}(N((K+1)M^2)^{3.5})$. 
Let $I_b$ denote the number of beamforming SCA 
iterations per AO iteration and $J$ the maximum 
number of convex subproblem solves per slot in each 
SCA iteration, including the multiplier search and 
auxiliary minimum-power computation. The resulting 
beamforming optimization complexity is 
$\mathcal{O}(I_bJN((K+1)M^2)^{3.5})$.
For the trajectory subproblem $\mathcal{P}7.l$, which 
jointly optimizes the flight speeds 
over $N$ time slots, the trust-region-based SCA 
formulation has a complexity of $\mathcal{O}(N^{3.5})$ per convex problem solve. Let $I_t$ 
denote the total number of trajectory subproblem 
solves per AO iteration, including those required 
for trust-region radius adjustments. Therefore, 
the overall per-AO-iteration 
computational complexity of the proposed algorithm 
can be approximated as $\mathcal{O}(I_bJN((K+1)M^2)^{3.5}
+I_tN^{3.5})$.

\par

\section{Proposed Algorithm for NTO}
\label{sec_method_NTO}
This section presents an AO algorithm to efficiently solve problem~$\mathcal{P}2$. The proposed approach iteratively updates the flight velocity sequence $\{V[n]\}_{n=1}^N$ and the HAPS altitude $H$ in an alternating manner to minimize the total propulsion power. Specifically, for a given altitude, the optimal velocity profile is first determined. Subsequently, the altitude is updated based on the obtained velocity, and this process is repeated until convergence.

\subsection{Velocity Optimization}
For a fixed altitude $H$, the corresponding air density can be obtained from the empirical model as $\rho_h=(0.95162H^2-52.29356H+753.39927)\times 10^{-3}(kg/m^3)$, then the flight velocity sequence $\{V[n]\}_{n=1}^N$ is optimized by solving
\begin{subequations}\label{P9}
\begin{align}
    \mathcal{P}8:&\mathop{\min}_{V[n],\forall n}\,\, \sum_{n=1}^{N}P_{\mathrm{SCF}}(V[n],H), \notag\\
    &\quad\,\mathrm{s.t.}\,\,\, (\ref{P2}b)\, \mathrm{and}\, (\ref{P2}c),   \notag
\end{align}
\end{subequations}
Since $P_{\mathrm{SCF}}(V[n],H)$ is convex with respect to $V[n]$ for all $n$, problem $\mathcal{P}8$ can be decomposed into $N$ independent convex subproblems, 
each minimizing $P_{\mathrm{SCF}}[n]$ individually. Consequently, the optimal velocity candidate at each time slot can be obtained by applying 
the first-order optimality condition, i.e., by setting the derivative of 
$P_{\mathrm{SCF}}[n]$ with respect to $V[n]$ to zero:
\begin{align}
    \frac{\partial P_{\mathrm{SCF}}(V[n],H)}{\partial V[n]} =& \frac{1}{\cos(\vartheta)^2f_pf_e}\Big(\frac{3}{2}\rho_hV[n]^2SC_{D_0}\notag\\
    &-\frac{2\epsilon F_{\omega}^2}{\rho_hSV[n]^2}\Big) \triangleq 0,
\end{align}
which yields
\begin{align}
    \tilde{V}[n] = \sqrt{\frac{2F_{\omega}}{\rho_hS}\sqrt{\frac{\epsilon}{3C_{D_0}}}}.
\end{align}
The optimal velocity at each time slot is then determined as
\begin{align}\label{optimal_V}
    V^{\star}[n] = \left\{\begin{array}{lll}
        \tilde{V}[n], & \tilde{V}[n] \,\,\mathrm{satisfys}\,(\ref{P2}b)\, \mathrm{and}\, (\ref{P2}c), \\
        V_{\mathrm{min}}, & \tilde{V}[n]\leq V_{\mathrm{min}}, \\
        V_{\mathrm{max}}, & \tilde{V}[n]\leq V_{\mathrm{max}}. 
    \end{array}\right.
\end{align}

\subsection{Altitude Optimization}
For a fixed velocity sequence ${V[n]}$, the altitude $H$ is optimized to minimize the total propulsion power. The corresponding optimization problem is formulated as
\begin{subequations}\label{P8}
\begin{align}
    \mathcal{P}9:&\mathop{\min}_{H}\,\, \sum_{n=1}^{N}P_{\mathrm{SCF}}(V[n],H), \notag\\
    &\,\,\,\mathrm{s.t.}\,\,\, (\ref{P2}a)\, \mathrm{and}\, (\ref{P2}c), \notag
\end{align}
\end{subequations}
Since problem $\mathcal{P}9$ contains only one optimization variable, $H$, and the propulsion power function $P_{\mathrm{SCF}}$ is convex with respect to $H$. Specifically, the gradient of $\sum_{n=1}^{N} P_{\mathrm{SCF}}(V[n],H)$ with respect to the HAPS altitude $H$ is expressed as
\begin{align}
    \frac{\partial \sum\limits_{n=1}^NP_{\mathrm{SCF}}(V[n],H)}{\partial H} =& \frac{1}{\cos(\vartheta)^2f_pf_e}\left(\frac{1}{2}\frac{\partial\rho_h}{\partial H}\sum_{n=1}^{N}V[n]^3SC_{D_0}\right. \notag\\
    &\left. - \frac{2\epsilon F_{\omega}^2}{\rho_h^2S}\sum_{n=1}^{N}\frac{1}{V[n]}\frac{\partial\rho_h}{\partial H}\right).
\end{align}
Setting this derivative to zero yields the air density at the stationary point $\tilde{H}$ as
\begin{align}
    \rho_h(\tilde{H}) = \frac{2F_{\omega}}{S}\sqrt{\frac{\epsilon\sum_{n=1}^N\frac{1}{V[n]}}{C_{D_0}\sum_{n=1}^NV[n]^3}}. 
\end{align}
Substituting the empirical air-density model $\rho_h(\tilde{H})=(0.95162\tilde{H}^2-52.29356\tilde{H}+753.39927)\times 10^{-3}(kg/m^3)$ into the above expression allows us to determine the corresponding candidate altitude $\tilde{H}$. The optimal altitude $H^{\star}$ is then given by
\begin{align}\label{optimal_H}
    H^{\star} = \left\{\begin{array}{lll}
        \tilde{H}, & \tilde{H}\,\, \text{satisfies}\,(\ref{P2}a)\, \mathrm{and}\, (\ref{P2}c), \\
        H_{\mathrm{min}}, & \tilde{H}\leq H_{\mathrm{min}}, \\
        H_{\mathrm{max}}, & \tilde{H}\geq H_{\mathrm{max}}.
    \end{array}\right.
\end{align}

\subsection{Convergence and Complexity Analysis}
The overall AO framework for solving problem $\mathcal{P}2$ is summarized in Algorithm~2 of the supplementary material. Since both the altitude and velocity subproblems admit closed-form optimal solutions, each iteration of Algorithm~2 strictly decreases the total propulsion power $\sum_{n=1}^{N}P_{\mathrm{SCF}}(V[n],H)$ until convergence. Hence, the proposed algorithm guarantees a monotonic decrease in the objective function and converges to a stationary point of problem~$\mathcal{P}2$. The computational complexity of each iteration is mainly determined by the evaluation of $P_{\mathrm{SCF}}(V[n],H)$ and its gradients, which scales linearly with the number of time slots, i.e., $\mathcal{O}(N)$. Therefore, the proposed NTO algorithm achieves fast convergence and maintains a very low computational overhead, making it well-suited for real-time implementation.

\begin{algorithm}[t]
\caption{NTO - Joint Altitude and Velocity Optimization}
\label{alg_NTO}
    \begin{algorithmic}[1]
        \REQUIRE $H_{\mathrm{min}}$, $H_{\mathrm{max}}$, $V_{\mathrm{min}}$, $V_{\mathrm{max}}$, $\Delta_t$, and $e_{\mathrm{start}}$.  
        \ENSURE $H^{\star}$ and $V^{\star}[n]$, $\forall n$. 
        \STATE Initialize $H^{(0)}$, and let $l=0$.  
        \WHILE{no convergence}
        \STATE Obtain $V^{(l+1)}[n]$ via (\ref{optimal_V}) for given $H^{(l)}$.  
        \STATE Obtain $H^{(l+1)}$ via (\ref{optimal_H}) for given $V^{(l)}[n]$, $\forall n$.
        \STATE Calculate $\sum_{n=1}^NP_{\mathrm{SCF}}[n]$ using $V^{(l)}[n]$ and $H^{(l)}$, $\forall n$. 
        \STATE Update $l=l+1$. 
        \ENDWHILE
        \STATE Return $H^{\star}=H$ and $V^{\star}[n]=V[n]$, $\forall n$. 
    \end{algorithmic}
\end{algorithm}

\par

\section{Simulation results}
\label{sec_sim}
In this section, numerical results are presented to evaluate the performance of the proposed algorithm. We consider a solar-powered HAPS-enabled ISAC system serving $K = 8$ users and sensing $Q = 8$ targets. The user and target locations are independently generated. Specifically, six nodes are uniformly generated within the inner coverage region defined by the SAR observation angle $\alpha = 2.5^\circ$, which may be interpreted as the dense urban coverage zone, while the remaining two nodes are uniformly placed within the larger region corresponding to $\beta = 5^\circ$, representing a broader open-area footprint such as suburban or desert environments. The HAPS is equipped with $M=12$ antennas. The carrier frequency is set to $f_0=9.6$ GHz, while the noise power gain at each user receiver is fixed at $\sigma_k^2=-60$ dBm. The beampattern gain threshold is set to $\Gamma=-40$\,dBm, and the maximum transmit power is fixed at $P_{\mathrm{max}}=10$\,W. The Rician factor is chosen as $K_u=10$. The total operation duration is $T=600~\mathrm{s}$, with a sampling interval of $\Delta_t=5~\mathrm{s}$, where the first circular segment spans $T_1=150~\mathrm{s}$. During the DTO phase, the HAPS flies at a fixed altitude of $H = 20~\mathrm{km}$, and its horizontal velocity is constrained within the range $V_{\min} = 10~\mathrm{m/s}$ to $V_{\max} = 40~\mathrm{m/s}$. Assuming that the HAPS operate over KAUST during the summer season, the harvested solar power is set to $P_a = 0.29\times 143\times 1.3 = 53.911$ kW in accordance with (20). During the NTO phase, the HAPS operates within the allowable altitude range, flying at a minimum altitude of $H_{\mathrm{min}}=15$ km and up to a maximum altitude of $H_{\mathrm{max}}=25$ km. 

\begin{figure}[t]
\centering
\includegraphics[width=0.73\linewidth]{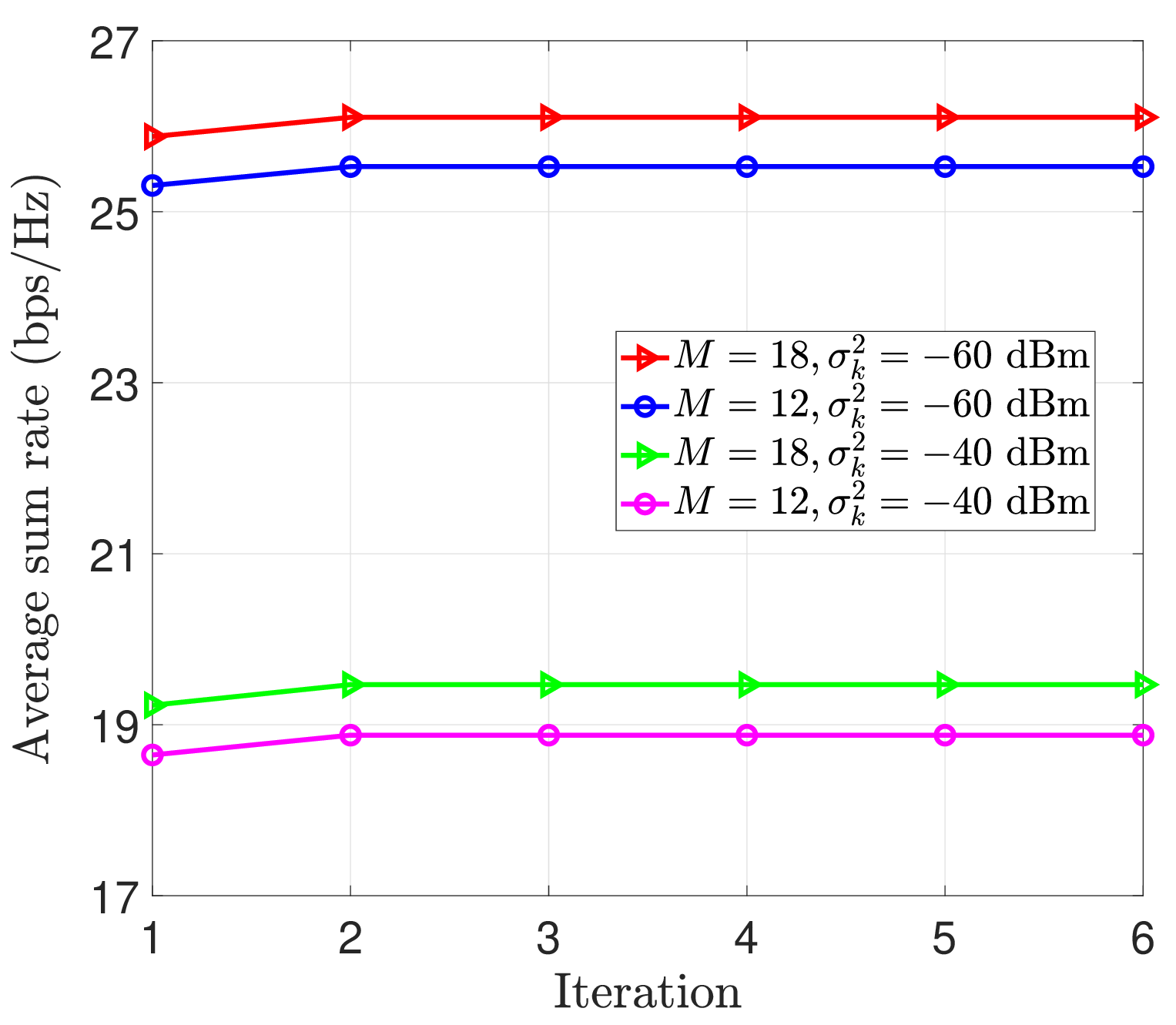}
\caption{Convergence of the proposed algorithm 1.}
\label{convergence_DTO}
\end{figure}

Before comparing system performance, we first examine the convergence behavior of Algorithm 1 for the DTO phase. As shown in Fig.~\ref{convergence_DTO}, the proposed algorithm exhibits a stable and monotonic improvement in sum-rate throughput across different antenna configurations $M$ and noise power levels $\sigma_v^2$. In all cases, the throughput converges rapidly within only a few iterations, confirming the fast and reliable convergence of the proposed optimization framework.

To evaluate the proposed framework, five benchmark schemes with different operational characteristics are considered during DTO. The large-circle trajectory enables continuous loitering over a single region for broad-area coverage, whereas the straight-flight trajectory facilitates point-to-point traversal with limited revisit capability. In contrast, the proposed figure-eight trajectory enables persistent ISAC services over two geographically separated regions through periodic revisits while maintaining continuous fixed-wing flight. The benchmark schemes are described as follows.
\begin{itemize}
    \item \textbf{Large-Circle Beamforming Design:} The HAPS follows a predetermined large-circle trajectory that fully encompasses the proposed figure-eight path. Similar to the proposed design, the flight speed profile $V[n]$, the communication beamforming vectors $\mathbf w_k[n]$, and the SAR covariance matrices $\mathbf R_s[n]$ are jointly optimized. The corresponding optimization problem is obtained from Problem~$\mathcal P1$ by replacing the figure-eight trajectory constraints with the large-circle trajectory constraints.
    \item \textbf{Straight-Flight Beamforming Design}: The HAPS follows a predetermined piecewise straight-flight trajectory. Starting from the initial position $\mathbf p[1]$, it first flies to $(0,-2R_1,H)$, then proceeds to the intersection point $\mathbf p[N_1]$, subsequently travels to $(0,2R_2,H)$, and finally reaches the terminal position $\mathbf p[N]$. Similar to the proposed design, the flight speed profile $V[n]$, the communication beamforming vectors $\mathbf w_k[n]$, and the SAR covariance matrices $\mathbf R_s[n]$ are jointly optimized. The corresponding optimization problem is obtained from Problem~$\mathcal P1$ by replacing the figure-eight trajectory constraints with the piecewise straight-flight trajectory constraints.
    \item \textbf{Constant-Speed Beamforming Design}: The HAPS follows the same figure-eight trajectory as the proposed design while maintaining a constant flight speed on each circular segment, i.e., $V[n]=\frac{2\pi R_1}{N_1\Delta t}$ for $n=1,\ldots,N_1$, and $V[n]=\frac{2\pi R_2}{(N-N_1)\Delta t}$ for $n=N_1+1,\ldots,N$. Compared with the proposed design, the flight speed profile is predetermined rather than optimized, whereas the communication beamforming vectors $\mathbf w_k[n]$ and the SAR covariance matrices $\mathbf R_s[n]$ are jointly optimized by solving Problem~$\mathcal P3$.
    \item \textbf{Isotropic Transmission Design}: The HAPS employs isotropic transmission, where the beamforming matrices are given by $\mathbf W_k[n]=\frac{P_{c,k}}{M}\mathbf I_M$ and $\mathbf R_s[n]=\frac{P_t}{M}\mathbf I_M$, $\forall k,n$, with $P_{c,k}$ and $P_t$ denoting the transmit powers allocated to the communication and dedicated SAR imaging signals, respectively. The corresponding total transmit power constraint is $\sum_{k=1}^{K}P_{c,k}+P_t\le P_{\max}$. Similar to the proposed design, the flight speed profile is jointly optimized with the transmit power allocation. The resulting optimization problem is obtained from Problem~$\mathcal P1$ by replacing the beamforming variables with the isotropic transmission model.
    \item \textbf{Communication-Only Beamforming Design}: The HAPS is dedicated exclusively to communication without performing SAR imaging. The flight speed profile $V[n]$ and the communication beamforming vectors $\mathbf w_k[n]$ are jointly optimized. The corresponding optimization problem is obtained from Problem~$\mathcal P1$ by removing the SAR-related variables and constraints, and is formulated as
    \begin{subequations}
    \begin{align}
    \mathcal{P}10: &\mathop{\max}\limits_{V[n]\atop \mathbf{w}_k[n],\forall k,n}\,\frac{1}{N}\sum_{n=1}^N\sum_{k=1}^K\mathcal{R}_k[n], \notag\\
    &\mathrm{s.t.}\,\sum_{k=1}^K\|\mathbf{w}_k[n]\|^2 \leq P_{\mathrm{max}},(\ref{P1}d)-(\ref{P1}f). \notag
\end{align}
\end{subequations}
Problem~$\mathcal P10$ can be solved by following the same AO-SCA procedure developed for Problem~$\mathcal P1$. The detailed derivation is omitted for brevity.
\end{itemize}

\begin{figure}[t]
\centering
\includegraphics[width=0.73\linewidth]{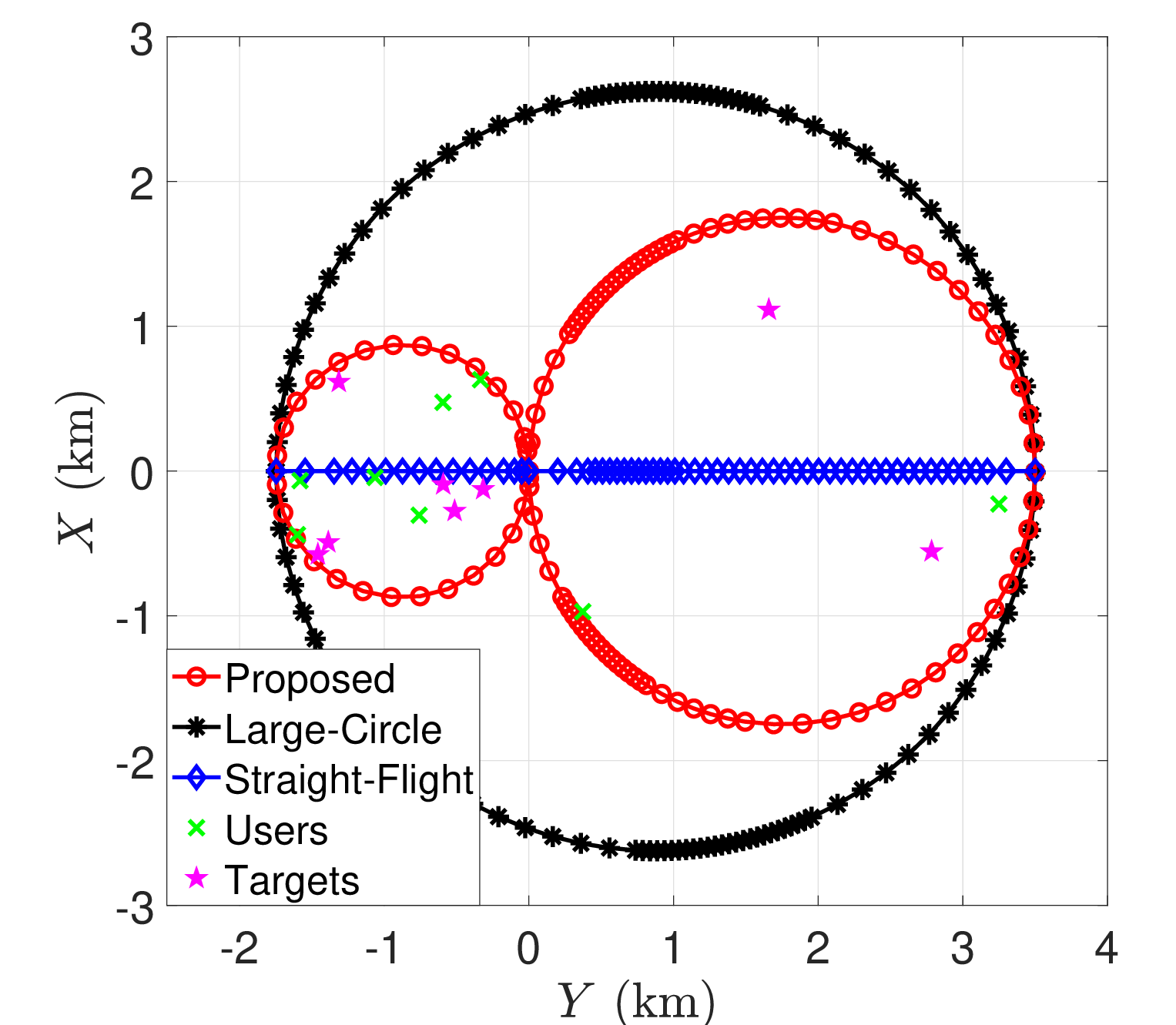}
\caption{Obtained trajectories of the HAPS during the DTO phase in the 2D horizontal plane.}
\label{optimal_trajectories_DTO}
\end{figure}

Fig.~\ref{optimal_trajectories_DTO} demonstrates the considered HAPS trajectories in the horizontal plane during the DTO phase. The straight-flight trajectory serves as a simple baseline, where the HAPS sequentially traverses the two service regions along piecewise linear flight segments. The large-circle trajectory continuously loiters over a single orbit but generally maintains larger communication and sensing distances to the users and targets. In contrast, the proposed figure-eight trajectory alternately loiters around the two service regions, allowing the HAPS to remain closer to each region for a longer duration during every flight cycle, thereby improving both communication and sensing performance. The Constant-Speed Beamforming and Isotropic Transmission benchmarks are omitted from Fig.~\ref{optimal_trajectories_DTO}, since both follow the same figure-eight trajectory as the proposed design.

\begin{figure}[t]
\centering
\includegraphics[width=0.73\linewidth]{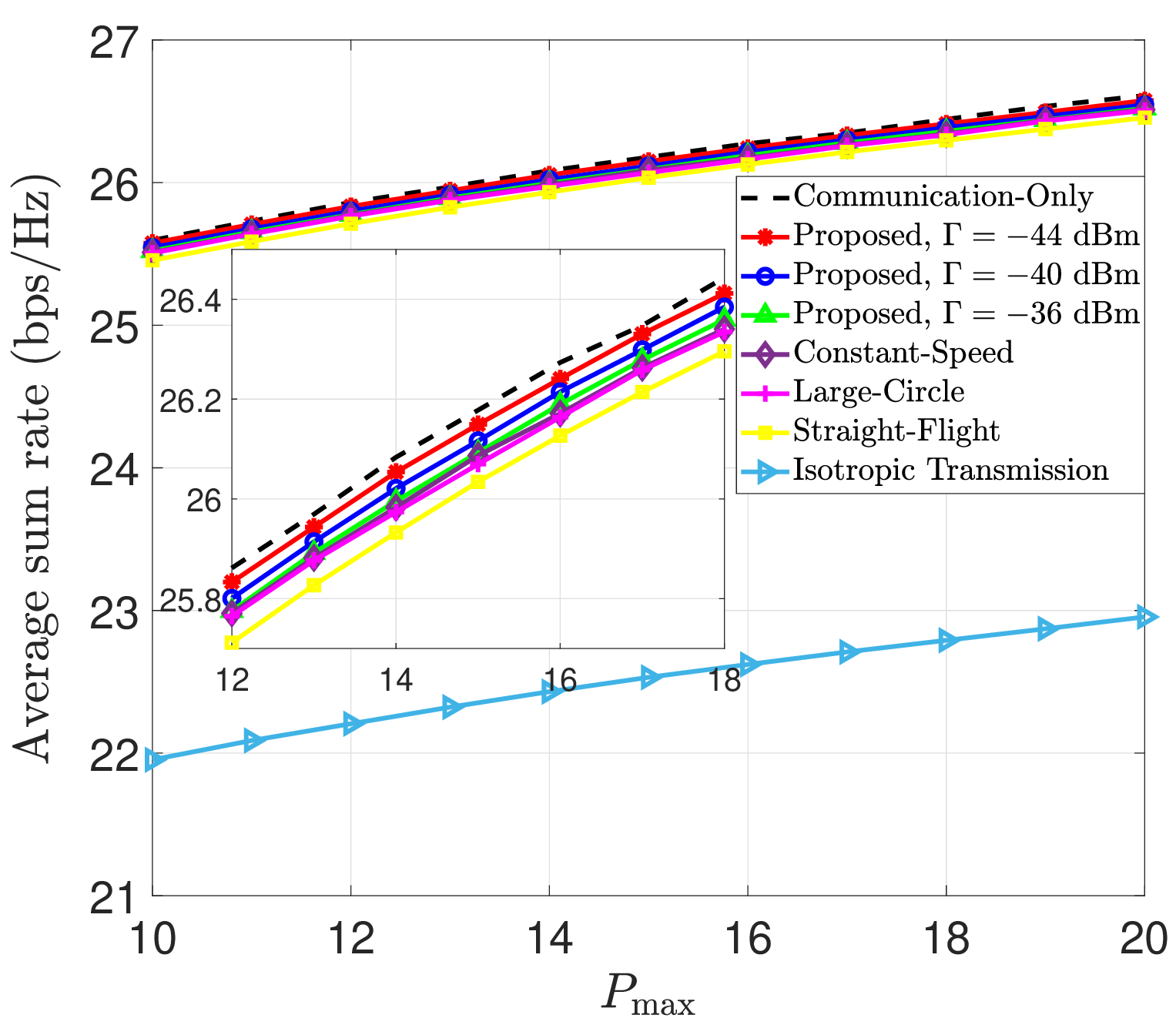}
\caption{The optimal average sum-rate throughput versus the maximum transmit power $P_{\mathrm{max}}$ under different $\Gamma$ for the DTO phase.}
\label{power_rate}
\end{figure}

\begin{figure}[t]
\centering
\includegraphics[width=0.735\linewidth]{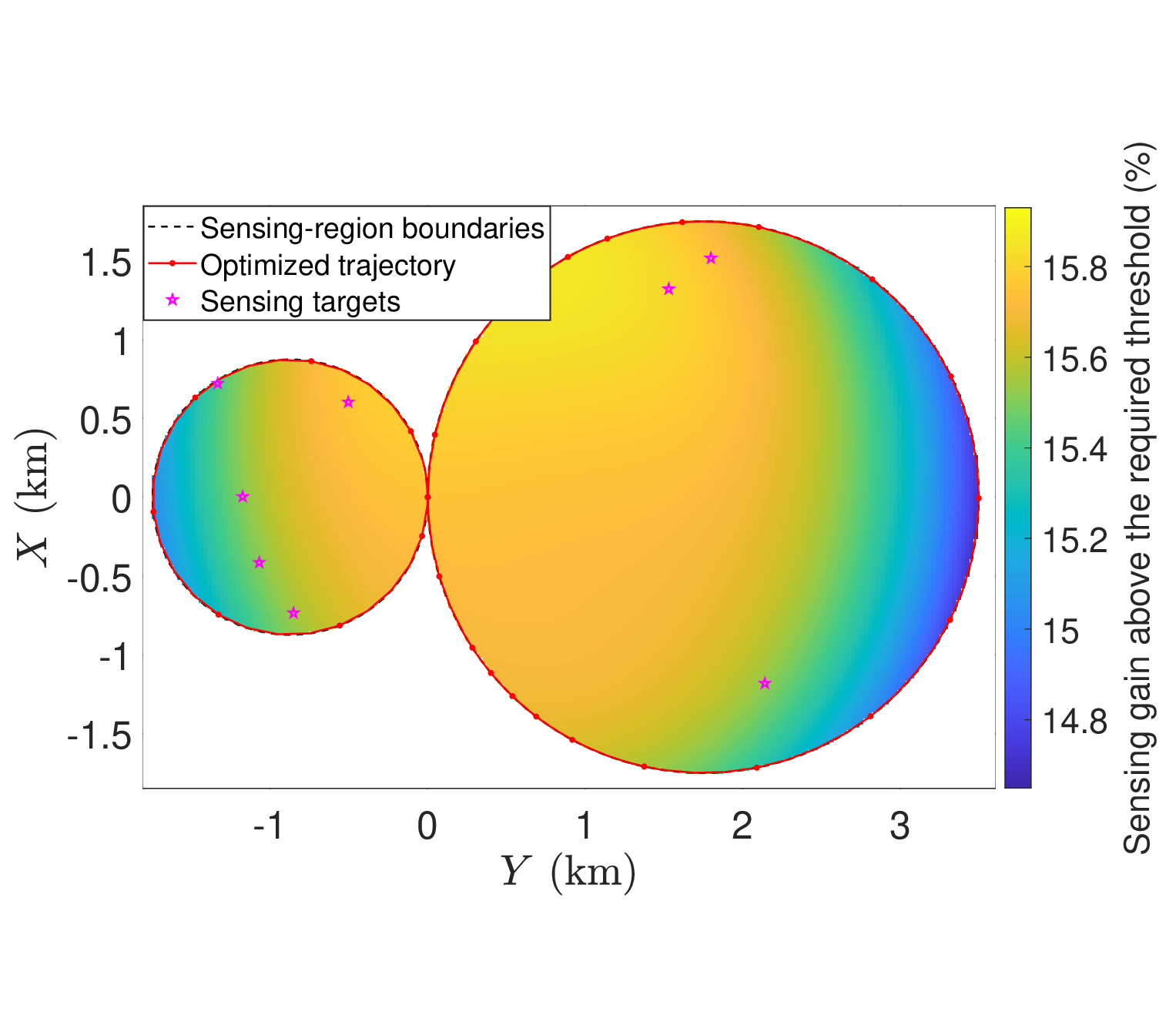}
\caption{Regional sensing performance of the proposed figure-eight trajectory. The color denotes the time-averaged normalized sensing gain.}
\label{sensing_region}
\end{figure}

\begin{figure}[t]
\centering
\includegraphics[width=0.73\linewidth]{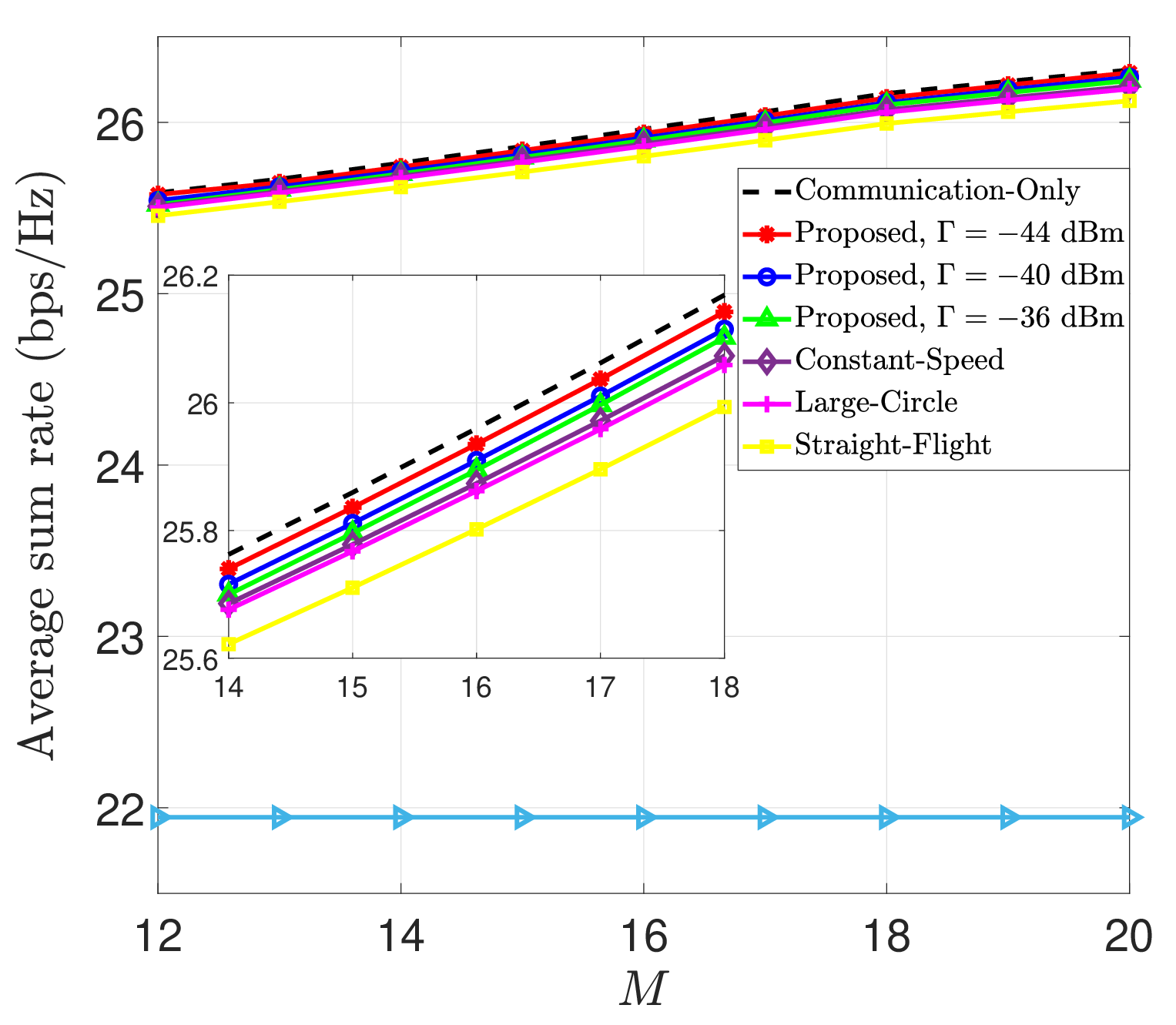}
\caption{The optimal average sum-rate throughput versus the number of antennas $M$ under different $\Gamma$ for the DTO phase.}
\label{M_rate}
\end{figure}

Additionally, Fig.~\ref{power_rate} illustrates the average sum-rate versus the maximum transmit power $P_{\mathrm{max}}$ under different sensing beampattern gain thresholds $\Gamma$. As expected, the sum-rate increases monotonically with $P_{\mathrm{max}}$ for all schemes. The proposed design consistently outperforms the Large-Circle, 
Straight-Flight, Constant-Speed Beamforming, and Isotropic Transmission benchmarks over the entire transmit-power range. The performance gains over the Constant-Speed Beamforming and Isotropic Transmission benchmarks demonstrate the benefits of flight-speed optimization and directional beamforming, respectively. As $\Gamma$ decreases, the relaxed sensing constraint provides greater beamforming flexibility, allowing the sum-rate to approach that of the Communication-Only benchmark. Notably, the proposed design maintains a small performance gap from the Communication-Only benchmark across the entire transmit-power range, indicating that the sensing requirements can be satisfied with only a marginal loss in communication performance.

To further evaluate spatial sensing coverage, Fig.~\ref{sensing_region} illustrates the regional sensing performance of the proposed figure-eight trajectory. A dense set of uniformly distributed test locations is generated within the two sensing regions, and the corresponding time-averaged sensing gain over one flight cycle is normalized by the required threshold $\Gamma$. The normalized sensing gain exceeds one at all test locations, indicating reliable coverage beyond the optimized sensing targets and yielding an empirical regional coverage probability of 100\%. Furthermore, the minimum normalized sensing gain of 1.147 corresponds to a sensing margin of 0.596\,dB, confirming reliable sensing coverage with a positive 
margin across both sampled regions.

Fig.~\ref{M_rate} illustrates the average sum-rate under different 
antenna array sizes $M$ and sensing beampattern gain thresholds 
$\Gamma$. As expected, the sum-rate increases with $M$ for all 
beamforming-based schemes owing to the enhanced beamforming capability and spatial degrees of freedom, whereas that of the Isotropic Transmission benchmark remains nearly unchanged. The proposed design consistently outperforms the Large-Circle, Straight-Flight, Constant-Speed Beamforming, and Isotropic Transmission benchmarks over the entire range of $M$. The performance gain over the Constant-Speed Beamforming benchmark confirms the effectiveness of joint flight-speed and transmit beamforming optimization across 
different array sizes, while the substantial gap from the Isotropic 
Transmission benchmark highlights the importance of directional 
beamforming. Notably, the proposed design achieves a sum-rate close 
to that of the Communication-Only benchmark across all considered 
array sizes, indicating a marginal communication performance loss 
due to the sensing requirements. This gap further narrows as 
$\Gamma$ decreases owing to the relaxed sensing constraint.

\begin{figure}[t]
\centering
\includegraphics[width=0.75\linewidth]{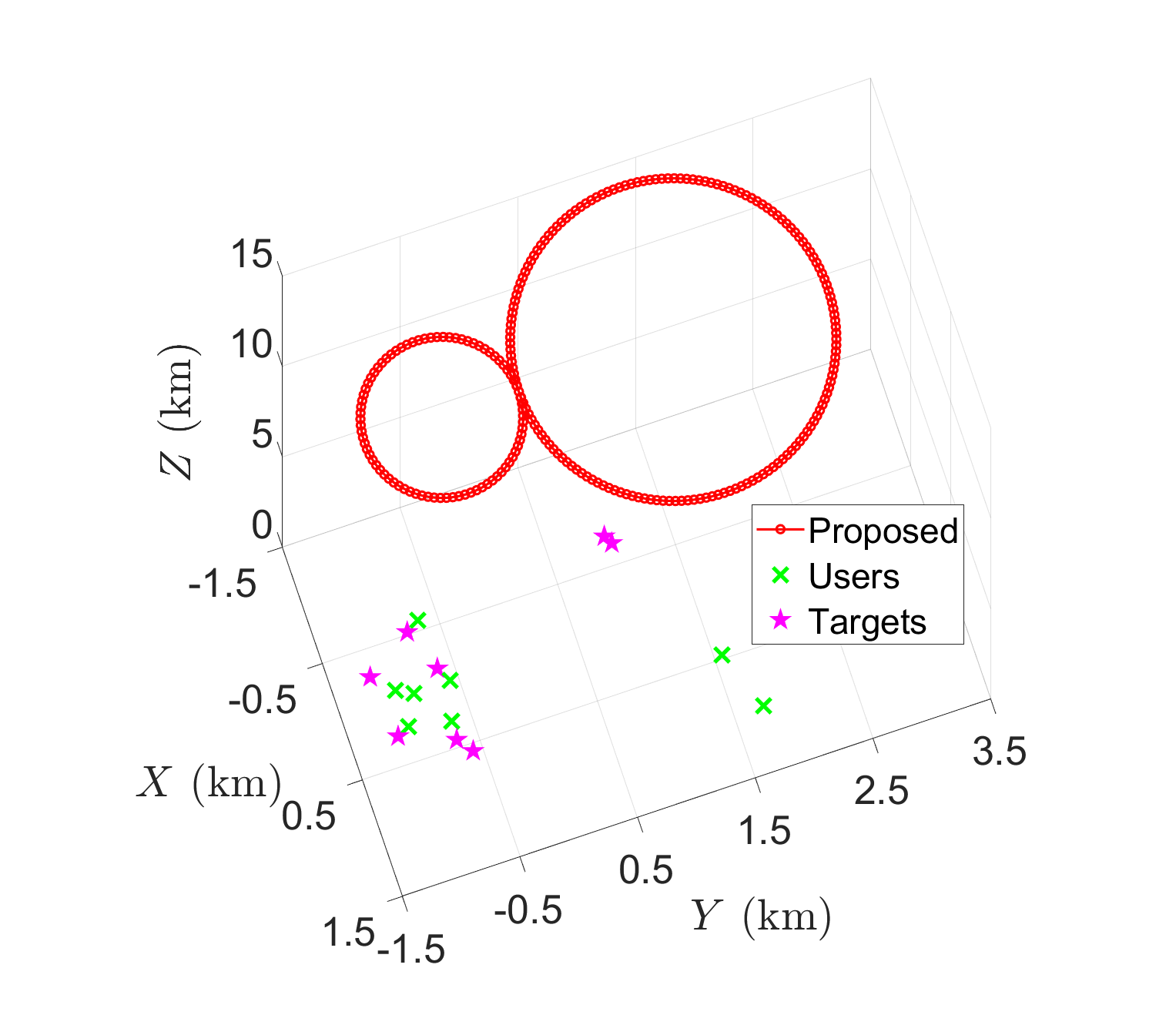}
\caption{Obtained trajectories of the HAPS during the NTO phase in 3D space.}
\label{optimal_trajectories_NTO}
\end{figure}

\begin{figure}[t]
\centering
\includegraphics[width=0.73\linewidth]{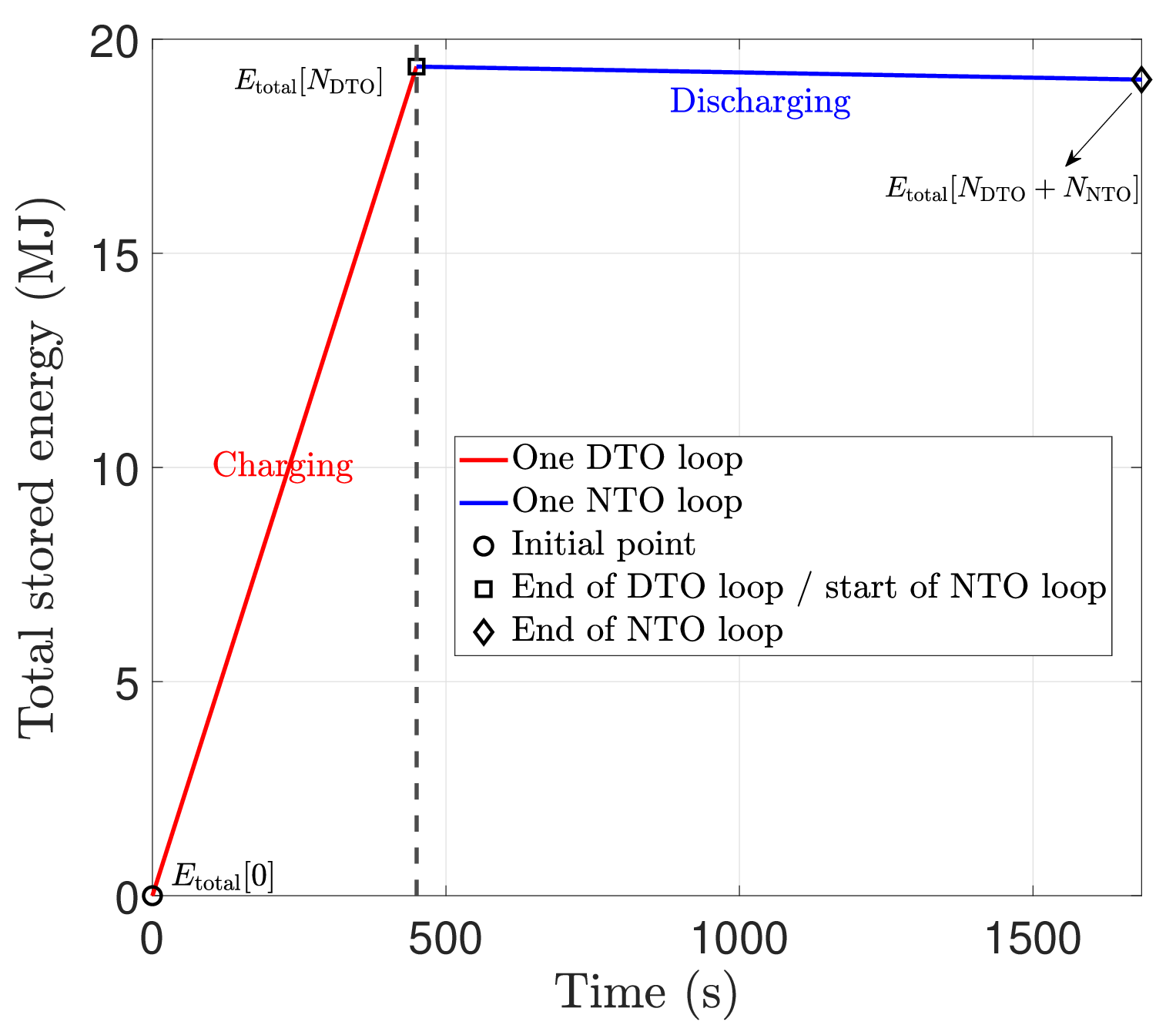}
\caption{Temporal evolution of the onboard energy over one DTO loop and one NTO loop along the figure-eight trajectory.}
\label{energy_Full_Day}
\end{figure}

Upon completion of DTO, the initial battery energy for NTO is given 
by $e_{\mathrm{start}}=E_{\mathrm{total}}[N+1]$. Fig.~\ref{optimal_trajectories_NTO} illustrates the energy-efficient nighttime figure-eight trajectory with the optimized altitude $H^{\star}=15$ km and flight speed $V^{\star}=10$ m/s. Compared with daytime operation, the reduced nighttime speed lowers propulsion power consumption and prolongs the trajectory traversal time while maintaining communication and sensing coverage for all users and targets. Fig.~\ref{energy_Full_Day} further illustrates the onboard battery energy evolution over one representative DTO loop and one NTO loop. During DTO, the battery energy increases continuously through solar energy harvesting, reaching $E_{\mathrm{total}}[N_{\mathrm{DTO}}]$ at the end of daytime operation. During NTO, the stored energy is gradually consumed to support propulsion and communication in the absence of solar power. 
Nevertheless, the remaining battery energy is positive at the end 
of the NTO loop, i.e., $E_{\mathrm{total}}[N_{\mathrm{DTO}}+N_{\mathrm{NTO}}]>0$, confirming the energy feasibility of the considered nighttime cycle. These results validate the proposed day-night energy management framework in coordinating solar energy harvesting and battery-powered 
operation to support sustainable long-endurance ISAC services.

\par

\section{Conclusion}
\label{sec_con}
This paper investigated a solar-powered HAPS-enabled ISAC framework that jointly integrates wireless communication, SAR imaging, and energy management for long-endurance aerial operation. A figure-eight loitering trajectory was proposed to provide efficient multi-region coverage, while a physics-based energy model was developed to characterize the coupled DTO and NTO phases. For the DTO phase, a joint beamforming and trajectory optimization problem was formulated to maximize the communication performance subject to SAR imaging, flight feasibility, and energy constraints, and was solved using a trust-region-assisted SCA algorithm. For the NTO phase, an energy-constrained propulsion power minimization problem was addressed through a low-complexity algorithm with closed-form updates for flight control. Simulation results demonstrated that the proposed framework achieves improved trajectory efficiency and communication–sensing trade-off performance while enabling sustainable day–night operation of solar-powered HAPS platforms.

\appendices

\bibliographystyle{IEEEtran}
\bibliography{references}

\end{document}